\documentclass[11pt]{article}       
\usepackage{geometry}               
\usepackage{hyperref}       
\usepackage{url}            
\usepackage{booktabs}       
\usepackage{amsfonts}       
\usepackage{xfrac}       
\usepackage{xcolor}         

\usepackage{graphicx} 
\usepackage{amsmath, amssymb}
\usepackage{mathtools}
\usepackage{mathrsfs}
\usepackage{amsthm}
\usepackage{bm}
\usepackage{bbm}
\usepackage[linesnumbered, ruled, vlined]{algorithm2e}
\usepackage{cleveref}
\usepackage{placeins}
\usepackage{cite}
\usepackage{multirow}
\usepackage{multicol}
\usepackage{array}
\usepackage{caption}
\usepackage{subcaption}
\usepackage{booktabs}
\usepackage{rotating}
\usepackage{makecell} 

\newtheorem{proposition}{Proposition}
\newtheorem{definition}{Definition}
\newtheorem{assumption}{Assumption}

\newtheorem{lemma}{Lemma}
\newtheorem{theorem}{Theorem}
\newtheorem{remark}{Remark}

\newcommand{\xpost}{\bar{x}}
\newcommand{\xprior}{\bar{x}^-}
\newcommand{\Xpost}{\Sigma_{x,t}}
\newcommand{\Xprior}{\Sigma_{x,t}^-}
\newcommand{\Xpostinit}{\Sigma_{x,0}}
\newcommand{\Xpriorinit}{\Sigma_{x,0}^-}
\newcommand{\Xpostnext}{\Sigma_{x,t+1}}
\newcommand{\Xpriornext}{\Sigma_{x,t+1}^-}
\newcommand{\Xpriorinitopt}{\Sigma_{x,0}^{-,*}}
\newcommand{\Xpostnextopt}{\Sigma_{x,t+1}^*}
\newcommand{\Xpriornextopt}{\Sigma_{x,t+1}^{-,*}}
\newcommand{\Xpostprev}{\Sigma_{x,t-1}}
\newcommand{\xnom}{\hat{x}_0^-}
\newcommand{\Xnom}{\hat{\Sigma}_{x,0}^-}
\newcommand{\Pdist}{\mathbb{P}}
\newcommand{\Qdist}{\mathbb{Q}}
\newcommand{\Qhat}{\hat{\mathbb{Q}}}
\newcommand{\ambset}{\mathcal{D}}
\newcommand{\Gauss}{\mathcal{N}}
\newcommand{\UQ}{\mathcal{UQ}}
\newcommand{\Priordist}{\Pdist_{x,t}^-}
\newcommand{\Priordistinit}{\Pdist_{x,0}^-}
\newcommand{\Postdist}{\Pdist_{x,t}}
\newcommand{\Postdistopt}{\Pdist_{x,t}^*}
\newcommand{\Priordistopt}{\Pdist_{x,t}^{-,*}}
\newcommand{\Priordistinitopt}{\Pdist_{x,0}^{-,*}}
\newcommand{\Postdistinit}{\Pdist_{x,0}}

\newcommand{\Wass}[2]{W_2(#1, #2)}
\newcommand{\Gelb}[2]{G(#1, #2)}
\newcommand{\real}[1]{\mathbb{R}^{#1}}
\newcommand{\symm}[1]{\mathbb{S}^{#1}}
\newcommand{\psd}[1]{\symm{#1}_{+}}
\newcommand{\pd}[1]{\symm{#1}_{++}}

\DeclareMathOperator{\Tr}{Tr}
\DeclareMathOperator*{\argmin}{arg\,min}

\graphicspath{ {./images/} }
\crefname{assumption}{assumption}{assumptions}

\SetCommentSty{mycommfont}

\title{Wasserstein Distributionally Robust Control and State Estimation for Partially Observable Linear Systems\thanks{This work was supported in part   by the Information and Communications Technology Planning and Evaluation (IITP) grant funded by MSIT(2022-0-00124, 2022-0-00480).}
}

\author{Minhyuk Jang\thanks{Department of Mechanical Engineering, Seoul National University, Seoul, South Korea {\tt\small jason4012@snu.ac.kr}}
\and Astghik Hakobyan\thanks{Center for Scientific Innovation and Education and National Polytechnic University of Armenia, Yerevan, Armenia {\tt\small astghikhakobyan@csie.am}}
 \and
 Insoon Yang\thanks{Department of Electrical and Computer Engineering and ASRI, Seoul National University, Seoul, South Korea {\tt\small insoonyang@snu.ac.kr}}
}

\date{}

\begin{document}

\maketitle

\begin{abstract}
This paper presents a novel Wasserstein distributionally robust control and state estimation algorithm for partially observable linear stochastic systems, where the probability distributions of disturbances and measurement noises are unknown. Our method consists of the control and state estimation phases to handle distributional ambiguities of system disturbances and measurement noises, respectively. Leveraging tools from modern distributionally robust optimization, we consider an approximation of the control problem with an arbitrary nominal distribution and derive its closed-form optimal solution. We show that the separation principle holds, thereby allowing the state estimator to be designed separately.  A novel distributionally robust Kalman filter is then proposed as an optimal solution to the state estimation problem with Gaussian nominal distributions. Our key contribution is the combination of distributionally robust control and state estimation into a unified algorithm. This is achieved by formulating a tractable semidefinite programming problem that iteratively determines the worst-case covariance matrices of all uncertainties, leading to a scalable and efficient algorithm. Our method is also shown to enjoy a guaranteed cost property as well as a probabilistic out-of-sample performance guarantee.
The results of our numerical experiments demonstrate the performance and computational efficiency of the proposed method. 
\end{abstract}

\section{Introduction}\label{sec:intro}
The linear-quadratic-Gaussian (LQG) control method, which is fundamental  in   control theory (e.g.,~\cite{kwakernaak1972linear}), handles partially observable linear systems with Gaussian uncertainties, employing the Kalman filter~\cite{simon2006optimal} for state estimation. 
However, such traditional control techniques rely heavily on the precise knowledge of uncertainty distributions, which is rarely available in complex real-world environments. For example, a robot navigating an unpredictable environment must make decisions using noisy sensor data. Instead of knowing the exact uncertainty distributions, it has only (poor) estimates derived from this imperfect data. In such cases, conventional control methods can fail, leading to catastrophic outcomes such as collisions.

To address this challenge, distributionally robust (DR) approaches to control and estimation have recently gained significant interest. These methods extend DR optimization (DRO) principles~\cite{chen2020distributionally, liu2022distributionally, lin2022distributionally} to develop robust and practical strategies for sequential decision-making. They aim to minimize expected costs under the worst-case distribution within an \emph{ambiguity set}, which includes distributions close to a nominal distribution constructed from data. Consequently, these methods effectively hedge against potential inaccuracies in the estimated distributional information.

In this work, we introduce a novel optimization-based framework for Wasserstein DR control and state estimation (WDR-CE) that simultaneously addresses distributional uncertainties in both control and estimation processes. Our approach concerns with finite-horizon linear-quadratic (LQ) settings in discrete time. Using techniques from DRO, particularly the Gelbrich bound on the Wasserstein distance~\cite{gelbrich1990formula}, we formulate an approximate penalty version of the Wasserstein DR control (DRC) problem, similar to \cite{hakobyan2024wasserstein}. Solving this yields a closed-form DRC policy that effectively manages ambiguity in the disturbance distribution. We establish the separation principle, allowing the independent design of a state estimator. This yields a separate Wasserstein DR state estimation (DRSE) problem to handle distributional errors in the initial state and measurement noise distributions. Its optimal solution is obtained as a novel DR Kalman filter, which significantly differs from existing approaches (e.g.,~\cite{nguyen2023bridging, NEURIPS_DRKF}). 

The core of our unified framework is a novel semidefinite programming (SDP) method, which consolidates the task of finding the worst-case covariance matrices for uncertainties in both control and estimation stages into a single problem. Our approach iteratively solves smaller SDP problems offline, enhancing computational efficiency and scalability. This results in a practical WDR-CE algorithm, with both the DR controller and state estimator derived in closed form, making it highly suitable for real-world applications.
A distinctive feature of our method is the affine structure of the worst-case mean of the disturbance distribution relative to the state estimates. This allows for updating the worst-case mean online and relaxes the zero-mean assumptions often used in the literature (e.g.,~\cite{taskesen2024distributionally}).

A guaranteed cost property is shown to hold, which demonstrates the distributional robustness of our controller.
 We also derive a probabilistic out-of-sample performance guarantee of our controller and state estimator.   The results of our numerical experiments validate the proposed method's capability to handle distributional ambiguities, including non-Gaussian and nonzero-mean distributions, and demonstrate its computational efficiency, particularly in solving problems over long horizons.
 
\subsection{Related work}
Our work focuses on sequential decision-making under imperfect distributions of underlying uncertainties. In this context, DRO has emerged as a powerful tool for providing solutions that are robust against distributional mismatches within a specified ambiguity set, typically constructed using a data-driven estimate~\cite{rahimian2022frameworks, lin2022distributionally, rahimian2019distributionally}.
Unlike robust optimization, DRO hedges against uncertainty in the probability distributions, balancing robustness and performance.  DRO frameworks define ambiguity sets using moment constraints~\cite{delage2010distributionally, nie2023distributionally, zymler2013distributionally}, relative entropy~\cite{ugrinovskii2002minimax, van2021data, li2021distributionally}, and $\phi$-divergences~\cite{ben2013robust, duchi2021learning}, among others. In particular, the Wasserstein ambiguity set has garnered significant attention~\cite{kuhn2019wasserstein,  inatsu2022bayesian,  chen2024data, gao2023distributionally, ning2023data, nietert2024outlier, gao2022wasserstein, azizian2023regularization, kannan2023residuals} for its tractability and out-of-sample performance guarantees~\cite{esfahani2015data, gao2023finite}. Wasserstein DRO has been successfully applied in various fields, especially in statistical and machine learning contexts~\cite{wu2024understanding, blanchet2019robust, blanchet2020semi, gao2017wasserstein, luo2019decomposition, shafieezadeh2015distributionally, chen2018robust, lee2018minimax, nguyen2022distributionally,huang2022coresets, shapiro2023bayesian,phan2023global, nguyen2024optimal, azizian2024exact}.

In control theory, various DRC methods have recently been developed exploiting techniques from modern DRO.\footnote{Another related line of researches focuses on distributionally robust (finite) Markov decision processes (MDPs)~\cite{xu2010distributionally,yu2015distributionally, yang2017convex,chen2019distributionally,clement2021first, nguyen2022distributionallycc} as well as distributionally robust reinforcement learning~\cite{yu2024fast, liu2022distributionally, xu2023group, shi2024curious, ramesh2024distributionally, abdullah2019wasserstein, queeney2024risk, kallus2022doubly, smirnova2019distributionally}.
However, distributionally robust formulations of partially observable MDPs (POMDPs)~\cite{nakao2021distributionally, saghafian2018ambiguous} are relatively sparse and subject to computational intractability in general.}
These methods have been explored in 
stochastic optimal control~\cite{van2015distributionally,yang2020wasserstein,kim2023distributional, taskesen2024distributionally, hakobyan2024wasserstein, al2023distributionally, hajar2023wasserstein, brouillon2023distributionally},
model predictive control~\cite{schuurmans2023safe, li2023distributionally, fochesato2022data, schuurmans2023general, kordabad2022safe, aolaritei2023wasserstein, zolanvari2023wasserstein, mcallister2023inherent, hakobyan2021wasserstein},
and data-driven control~\cite{coulson2021distributionally, mark2021data, navsalkar2023data, micheli2022data, liu2023data}. 
In particular,~\cite{hakobyan2024wasserstein} is closely related to our method in terms of the approximation technique used in controller design. However, it assumes a known initial state and measurement noise distributions and uses the standard Kalman filter for state estimation. 
Another related method is the finite-horizon DR-LQG control method proposed in~\cite{taskesen2024distributionally}. However, it is built upon the restrictive assumption that the nominal distributions of all uncertainties are zero-mean Gaussian, and requires a solution to a large SDP problem, making its scalability questionable.
In contrast, our method does not assume zero-mean distributions of uncertainties. Instead, we iteratively solve small SDP problems using off-the-shelf solvers, significantly improving scalability and practical applicability.

Concurrently, DRSE has evolved to address distributional errors in the state estimation, enhancing robustness against inaccuracies in prior state and measurement noise distributions~\cite{xie2023adjusted, wang2021distributionally, wang2022distributionally, wang2021robust, brouillon2023regularization, han2023distributionally, lotidis2023wasserstein, aolariteiwasserstein}. Significant developments include the Wasserstein DR Kalman filter~\cite{NEURIPS_DRKF} and the DR minimum mean square error (DR-MMSE) estimator~\cite{nguyen2023bridging}. However, integrating DRC and DRSE to address distributional uncertainties in both system dynamics and measurement process remains relatively unexplored.

\section{Problem setup} \label{sec:prelimi}

Consider a discrete-time linear stochastic system
\begin{equation}\label{eqn:system eq}
\begin{split}
    x_{t+1} &= A x_{t} + B u_{t} + w_t,\\
     y_t &= C x_t + v_t,
\end{split}
\end{equation}
where $x_t\in\real{n_x}, u_t\in\real{n_u}$ and $y_t \in\real{n_y}$ represent the system state, control input,  and output, respectively. The system disturbance $w_t \in \real{n_x}$ and measurement noise $v_t \in \real{n_y}$ are governed by distributions $\Qdist_{w,t} \in \mathcal{P}(\real{n_x})$ and $\Qdist_{v,t} \in \mathcal{P}(\real{n_y})$, respectively, while the initial state follows  $\Qdist_{x,0} \in \mathcal{P}(\real{n_x})$, where $\mathcal{P}(\mathcal{W})$ denote the set of Borel probability measures with support $\mathcal{W}$. Furthermore, the random vectors $w_t$, $v_t$, and the initial state $x_0$ are mutually independent, while $w_t$ and $w_{t'}$, as well as $v_t$ and $v_{t'}$, are independent for $t \neq t'$. 

The objective is to design an optimal controller $\pi:=(\pi_0, \dots, \pi_{T-1})$ for the system over a fixed horizon $T > 0$, minimizing the expected cumulative cost 
\[
\mathbb{E}_{x_0, v_t, w_t \forall{t}}\left[x_T^\top Q_f x_T + \sum_{t=0}^{T-1} x_t^\top Q x_t + u_t^\top R u_t\right],
\]
where $Q, Q_f\in\psd{n_x}$, $R\in\pd{n_u}$ are some weighting matrices.\footnote{Here, $\psd{n}$( $\pd{n}$) represents the cone of symmetric positive semi-definite (positive definite) matrices in $\symm{n}$. For any $A, B \in \symm{n}$, the relation $A\succeq B$ ($A\succ B$) means that $A-B\in\psd{n}$ $(A-B\in\pd{n})$.} 
However, as we are dealing with a partially observable system, at each time stage $t$, the control input is based on the information vector defined as $I_t \coloneqq (y_0,\dots,y_t,u_0,\dots,u_{t-1}) \; \forall t > 0$, with $I_0 \coloneqq y_0$.
In addition to the difficulty from partial observability, the distributions of the uncertainties acting on the system are unknown, with only nominal estimates $\Qhat_{x,0}, \Qhat_{w,t}$, and $\Qhat_{v,t}$ available. These nominal distributions can be chosen as empirical distributions or other estimates. These estimates are inherently unreliable for designing an optimal controller, particularly in partially observable environments. This adds complexity to the problem, as it necessitates the design of an optimal state estimator that can effectively handle the uncertainty in these distributions.

To address these challenges, we explore two distinct problems: a DRC problem that handles ambiguities in system disturbance distributions with fixed state estimator, initial state, and measurement noise distributions; and a DRSE problem that hedges against uncertainties in initial state and measurement noise distributions given a fixed prior state distribution.
While DRC and DRSE manage different uncertainties, we propose a unified solution to address all distributional ambiguities.

\textbf{Notation.} Given distributions $\Pdist \in \mathcal{P}(\mathcal{W}_1)$ and $\Qdist \in \mathcal{P}(\mathcal{W}_2)$, let $\Pdist \times \Qdist$ denote the product measure on the product space $\mathcal{W}_1 \times \mathcal{W}_2$.
Let $\Priordist$ and $\Postdist$ denote the prior and posterior distributions of the state $x_t$ at time $t$ given information $I_{t-1}$ and $I_t$, respectively. Define the conditional mean vector and covariance matrix of state $x_t$ corresponding to $\Priordist$ as $\xprior_t \coloneqq \mathbb{E}_{x_t}[ x_t \, \vert \, I_{t-1}]$ and $\Xprior \coloneqq \mathbb{E}_{x_t}[ (x_t - \xprior_t)(x_t - \xprior_t)^\top \, \vert \, I_{t-1}]$. Similarly, define the conditional mean vector and covariance matrix of state $x_t$ corresponding to $\Postdist$ as $\xpost_t \coloneqq \mathbb{E}_{x_t}[ x_t \, \vert \, I_t]$ and $\Xpost \coloneqq \mathbb{E}_{x_t}[ (x_t - \xpost_t)(x_t - \xpost_t)^\top \, \vert \, I_t]$.

\subsection{DRC problem}\label{sec:DRC}

Consider an adversarial policy $\gamma := (\gamma_0, \ldots, \gamma_{T-1})$ which at each time $t$ maps the information vector $I_t$ to the distribution of system disturbances from some ambiguity set $\ambset_{w,t} \subset \mathcal{P}(\mathbb{R}^{n_x})$ to be defined later. Define an auxiliary distribution $\Qdist_e:= (\Qdist_{e,0},\dots, \Qdist_{e,T-1})$ that combines the initial state and noise distributions as follows: $\Qdist_{e,0}:= \Qdist_{v,0} \times \Qdist_{x,0}$ and $\Qdist_{e,t}:= \Qdist_{v,t}$ at $t=0$ and $t>0$, respectively. Also, suppose a state estimator $\phi:=(\phi_0,\dots,\phi_{T-1})$ is given.
Then, for a fixed pair $(\phi, \Qdist_e)$, the DRC problem can be formulated as the following minimax control problem:
\begin{align}\label{eqn:minimaxbegin}
    \begin{split}
        \min_{\pi \in \Pi}\max_{\gamma \in \Gamma_{\ambset}}J_c(\pi, \gamma, \phi,\Qdist_e),
    \end{split}
\end{align}
where  the controller aims to
minimize the cost, while the adversary seeks to maximize it. Here, $\Pi = \{\pi:=(\pi_0,\dots,\pi_{T-1}) \mid \pi_t(I_t) = u_t\in\real{n_u},  \pi_t \mbox{ is measurable } \forall t\}$, and
$\Gamma_\ambset = \{\gamma:=(\gamma_0,\dots,\gamma_{T-1}) \mid \gamma_{t}(I_{t}) = \Pdist_{w,t} \in \ambset_{w,t}, \gamma_t \mbox{ is measurable } \forall t \}$ are the spaces of admissible control and disturbance distribution policies, while the cost function is given by
\[
J_c(\pi, \gamma, \phi,\Qdist_e)\coloneqq \mathbb{E}_\mathbf{y}\bigg[ \mathbb{E}_{x_T}\big[x^{\top}_T Q_f x_T \mid I_{T-1}\big] + \sum_{t=0}^{T-1}\mathbb{E}_{x_t}\big[x_t^{\top}Qx_t + u_t^{\top}Ru_t \mid I_t,u_t\big] \bigg],
\]
where $\mathbf{y}:=(y_0,\dots,y_T)$. Solving~\eqref{eqn:minimaxbegin} results in optimal control policy $\pi^*$ that is robust against inaccuracies in the nominal system disturbance distribution.

\subsection{DRSE problem}\label{sec:DRSE}

Since state estimation is an online process, it is often more convenient to consider its recursive version. Specifically, suppose at time $t$, we are given a prior state distribution $\Priordist$. Then, the goal is to design an estimator $\phi_t$ that maps the measurement $y_t$ to an optimal state estimate by minimizing the following estimation error:
\[
J_{e,t}(\phi_t, \Qdist_{v,t}, \Priordist): = \mathbb{E}_{x_t, v_t} \left[ \Vert x_{t}-\phi_{t}(Cx_t+v_t)\Vert_{\Theta_t} ^2 \mid I_{t-1}\right],
\]
where $\Theta_t \in \psd{n_x}$ is some weighting matrix. Given the distributions $\Qdist_{v,t}$ and $\Priordist$, the optimal state estimation problem constitutes a weighted MMSE estimation problem.
However, since we are given only nominal distributions of the measurement noise and initial state, there is a need to design a state estimator that hedges against uncertainties in these distributions.

This motivates us to consider a minimax optimization problem defined in each stage, where the state estimator aims to minimize the estimation error, while the measurement noise and initial state distributions $\Pdist_{v,t}$ and $\Priordistinit$ are chosen from respective ambiguity sets $\ambset_{v,t} \subset \mathcal{P}(\real{n_y})$ and
$\ambset_{x,0} \subset \mathcal{P}(\real{n_x})$ to maximize this error. 
Specifically, we define the DRSE problem at time $t$ as
\begin{equation} \label{eqn:DRSE}
\begin{split}
    &\min_{\phi_{t} \in \mathcal{F}_t} \max_{\Pdist_{e,t} \in \ambset_{e, t} } J_{e,t}(\phi_t, \Pdist_{v,t}, \Priordist),
    \end{split}
\end{equation}
where $\mathcal{F}_t = \{ \phi_t \mid \phi_t(y_t) \in \real{n_x},  \phi_t \mbox{ is measurable}\}$ denotes the space of admissible estimators, while the ambiguity set $\ambset_{e,t}$ is defined as $\ambset_{e,0} := \ambset_{v,0} \times \ambset_{x,0}$ and $\ambset_{e,t} := \ambset_{v,t}$  at $t=0$ and $t > 0$, respectively. The auxiliary distribution $\Pdist_{e,t}$ is defined as $\Pdist_{e,0}:=\Pdist_{v,0} \times \Priordistinit$ and $\Pdist_{e,t}:=\Pdist_{v,t}$ at $t=0$ and $t > 0$, respectively.\footnote{Unlike conventional state estimation problems, which typically minimize the squared error, our approach incorporates the weighting matrix $\Theta_t$, which allows for a more tailored and effective estimation process and is driven by the form of our DRC problem.} 

Our formulation significantly differs from prior DRSE methods (e.g., \cite{NEURIPS_DRKF, nguyen2023bridging}) as we are focusing solely on errors in measurement noise and initial state distributions. At the initial time stage, we address ambiguities in both the initial state and measurement noise distributions, then shift the focus to errors in the measurement noise distribution in subsequent stages. Our approach avoids unnecessary conservativeness of existing methods
 due to their independent treatment of uncertainties in the control and estimation phases 
 (See \Cref{app:DRSE_compare} for detailed comparisons).

\subsection{Wasserstein ambiguity set and Gelbrich distance} \label{sec:Wass_and_Gelbr}

In this work, we focus on Wasserstein ambiguity sets for both DRC and DRSE problems. 
We specifically employ the type-2 Wasserstein distance to measure the distance between distributions.

\begin{definition}[Wasserstein Distance]
    The type-2 Wasserstein distance between two probability distributions $\Pdist\in \mathcal{P}(\mathcal{W})$ and $\Qdist \in \mathcal{P}(\mathcal{W})$ is defined as:
\begin{align}
    W_2(\Pdist, \Qdist) \coloneqq \inf_{\tau \in \mathcal{T}(\Pdist, \Qdist)} \biggl\{ \biggl(\int_{\mathcal{W} \times \mathcal{W}} {\|x-y\|}^2 d\tau(x,y) \biggr)^{\frac{1}{2}} \biggr\}, \nonumber
\end{align}
where $\| \cdot \|$ is an arbitrary norm defined on $\mathcal{W}$, and $\mathcal{T}(\Pdist, \Qdist)$ is the set of joint probability distributions on $\mathcal{W} \times \mathcal{W}$ with marginals $\Pdist$ and $\Qdist$ respectively.
\end{definition}

The Wasserstein ambiguity sets in our work are defined as statistical balls centered on the respective nominal distributions, with the distance between two distributions measured using the type-2 Wasserstein distance. These sets are constructed as follows:
\begin{align*}
    \ambset_{w,t} &\coloneqq \big\{ \Pdist_{w,t} \in \mathcal{P}(\real{n_x}) \mid \Wass {\Pdist_{w,t}}{\Qhat_{w,t}} \leq \theta_{w} \big\}\\
        \ambset_{v,t} &\coloneqq \big\{ \Pdist_{v,t} \in \mathcal{P}(\real{n_y}) \mid  \Wass{\Pdist_{v,t}}{\Qhat_{v,t}} \leq \theta_{v} \big\}\\        \ambset_{x,0} &\coloneqq \big\{ \Priordistinit \in \mathcal{P}(\real{n_x}) \mid  \Wass{\Priordistinit}{\Qhat_{x,0}^-} \leq \theta_{x_0} \big\},
\end{align*}
where $\theta_w, \theta_{v}, \theta_{x_0} \geq0$ are the radii of the corresponding balls, determining the level of robustness against the nominal distributions $\Qhat_{w,t}, \Qhat_{v,t}$, and $\Qhat_{x,0}^-$. A radius of zero indicates a strong belief in the nominal distribution.

However, measuring the Wasserstein distance between two distributions directly is often infeasible, especially with partial observations, as highlighted in~\cite[Appendix A]{hakobyan2022wasserstein}. 
To overcome this challenge, we turn to the Gelbrich distance, which provides a practical alternative.

\begin{definition}[Gelbrich Bound]
    The Gelbrich distance between two distributions $\Pdist \in \mathcal{P}(\mathbb{R}^n)$ and $\Qdist \in \mathcal{P}(\mathbb{R}^n)$ with mean vectors $\mu_1, \mu_2 \in \real{n}$ and covariance matrices $\Sigma_1, \Sigma_2\in \psd{n}$ is defined as:
    \[
     \Gelb{\Pdist}{\Qdist} \coloneqq \sqrt{\|\mu_1-\mu_2\|_2^2 + B^2(\Sigma_1, \Sigma_2)}, 
    \]
    where $B^2(\Sigma_1, \Sigma_2) \coloneqq \Tr[\Sigma_1+\Sigma_2-2\big(\Sigma_2^{\frac{1}{2}} \Sigma_1 \Sigma_2^{\frac{1}{2}}\big)^{\frac{1}{2}}]$ is the squared Bures–Wasserstein distance. 
\end{definition}
The Gelbrich distance provides a lower bound for the type-2 Wasserstein distance  with the Euclidean norm, i.e., $\Gelb{\Pdist}{\Qdist} \leq W_2(\Pdist, \Qdist)$, with equality for elliptical distributions having the same density generators~\cite{gelbrich1990formula}. Utilizing only first and second-order moments and involving algebraic operations, the Gelbrich distance plays an important role in converting the DRC problem into a tractable form. 

\section{Tractable solutions to DRC and DRSE problems}\label{sec:Trac_approx_and_sol}

In this section, we present our WDR-CE framework for addressing distributional uncertainties. First, we approximately solve the DRC problem, then introduce a DR Kalman filter for the DRSE problem. Finally, we combine these methods into a unified, tractable, and scalable WDR-CE algorithm.

\subsection{Approximate DRC problem and its solution} \label{sec:approxDRC}

Directly solving the DRC problem~\eqref{eqn:minimaxbegin} is challenging due to its infinite-dimensionality and partial observability. We adopt the approximation technique from \cite{hakobyan2024wasserstein}, replacing the Wasserstein constraint with a Gelbrich distance penalty term. This approximation reformulates the DRC problem as follows:
\begin{align}\label{eqn:minimax}
    \begin{split}
        \min_{\pi \in \Pi}\max_{\gamma \in \Gamma}J_c^\lambda(\pi, \gamma, \phi,\Qdist_e)
    \end{split}
\end{align}
where the cost function is defined as
\[
J_c^{\lambda}(\pi, \gamma) := J_c (\pi, \gamma, \phi,\Qdist_e) - \lambda \sum_{t=0}^{T-1} \Gelb{\Pdist_{w,t}}{\Qhat_{w,t}}^2,
\]
which integrates the standard LQ objective $J_c$ with a penalty for deviation from the nominal distribution. 
Here, the space of admissible distribution policies is defined as $\Gamma = \{\gamma:=(\gamma_0,\dots,\gamma_{T-1}) \mid \gamma_{t}(I_{t}) = \Pdist_{w,t} \in \mathcal{P}(\real{n_x}) , \; \gamma_t \mbox{ is measurable } \forall t\}$, which is less restrictive than $\Gamma_\ambset$, as it does not constrain $\Pdist_{w,t}$ to be selected from $\ambset_{w,t}$.
Instead, the deviations from the nominal distribution $\Qhat_{w,t}$ are penalized by the squared Gelbrich distance, while the penalty parameter $\lambda >0$ balances robustness against worst-case scenarios with a preference for distributions close to the nominal one. 

Due to the properties of the Gelbrich bound detailed in~\Cref{sec:Wass_and_Gelbr}, this approximation technique is appealing for its tractability, resulting in a closed-form expression for the optimal control policy and worst-case system disturbance mean, formalized in the following theorem. Additionally, it provides a guarantee on the original cost function, allowing for the selection of the optimal parameter $\lambda$ to minimize the optimality gap, as discussed in~\Cref{sec:properties}. 

Let $\Phi \coloneqq BR^{-1}B^{\top} - \lambda^{-1} I \in \symm{n_x}$ and consider coefficients $P_{t}\in\psd{n_x}, S_{t}\in\psd{n_x}, r_{t}\in\real{n_x}$, and $q_{t}\in\real{}$
determined recursively through the Riccati equation in~\Cref{app:sol}.

\begin{assumption} \label{assump:lambda}
    The penalty parameter satisfies $\lambda I \succ P_t$ for all $t=1,\dots,T$.
\end{assumption}
\begin{theorem}{\cite[Theorem 1]{hakobyan2024wasserstein}} \label{thm:sol}
Suppose~\Cref{assump:lambda} holds and let $\hat{w}_t\in\real{n_x}$ and $\hat{\Sigma}_{w,t}\in\psd{n_x}$ denote the mean vector and covariance matrix of $w_t$ under $\Qhat_{w,t}$, respectively. Then, the optimal value to~\eqref{eqn:minimax} is given by
\begin{equation}\label{opt_val}
J_c^{\lambda}(\pi^*,\gamma^*, \phi,\Qdist_e) = \mathbb{E}_{x_0}[x_0^\top P_0 x_0 + \xi_0^\top S_{0}\xi_0  + 2 r_0^\top x_0] + q_{0} + \mathbb{E}_{\mathbf{y}}\left[\sum_{t=0}^{T-1} z_t(I_t) \right],
\end{equation}
where $z_t (I_t)$ for $t=0,\dots, T-1$ is given by
\begin{align}\label{max_sigma}
 z_t(I_t) := \sup_{\Sigma_{w,t} \in \psd{n_x} } \mathbb{E}_{y_{t+1}}\Big[ \mathrm{Tr}\big[S_{t+1} \Xpostnext + (P_{t+1} - \lambda I )\Sigma_{w,t}+ 2 \lambda \big(\hat{\Sigma}_{w,t}^\frac{1}{2} \Sigma_{w,t} \hat{\Sigma}_{w,t}^\frac{1}{2}\big)^{\frac{1}{2}}\big] \; \Big\vert \;I_{t} \Big].
\end{align}
Moreover, if~\eqref{max_sigma} attains an optimal solution, then an optimal policy pair $(\pi^*,\gamma^*)$ can be obtain as follows:
The optimal control policy at each time $t$ is uniquely given by
\begin{align}\label{contr_param}
    \pi^*_t(I_t) =  K_t \xpost_t + L_t,
\end{align}
where 
 \begin{equation}\label{KL}
 \begin{split}
 K_t&:=-R^{-1}B^\top(I+P_{t+1}\Phi)^{-1} P_{t+1} A\\
 L_t&:=-R^{-1}B^\top(I+P_{t+1}\Phi)^{-1}( P_{t+1} \hat{w}_{t} + r_{t+1}).
 \end{split}
 \end{equation}

Consider a worst-case distribution $\Pdist_{w,t}^{*}$ characterized by a mean vector $\bar{w}_{t}^* \in \real{n_x}$ and a covariance matrix $\Sigma_{w,t}^* \in \psd{n_x}$, where the mean vector is uniquely defined as
\begin{equation}\label{mu}
    \bar{w}_{t}^*  = H_t \xpost_t + G_t,
\end{equation}
with 
 \begin{equation}\label{HG}
     \begin{split}
         H_t &:= (\lambda I - P_{t+1})^{-1} P_{t+1} (A + B K_t)\\
         G_t &: = (\lambda I - P_{t+1})^{-1} (P_{t+1} B L_t + r_{t+1} + \lambda \hat{w}_t),
     \end{split}
 \end{equation}
while the covariance matrix is obtained as a maximizer $\Sigma_{w,t}^{*}$ of the right-hand side of~\eqref{max_sigma}. Then, at each time $t$, $\gamma_t^*(I_t)=\Pdist_{w,t}^*$ is an optimal policy for the adversary.
\end{theorem}
Details can be found in~\cite{hakobyan2024wasserstein} and \Cref{app:sol}. 
From~\eqref{contr_param}, we observe that the DR control policy $\pi_t^*$ is affine in the conditional state mean $\xpost_t$, requiring state estimation under the worst-case distribution $\Pdist_{w,t}^*$. Also, \Cref{thm:sol} shows that the control policy $\pi^*_t$ is independent of the available information, resulting in the \textit{separation principle}, allowing the
control policy to be designed independently from the state estimator.

\begin{remark}\label{cor:separ_se_under_wc}
     A key aspect of our approach is that 
the conditional state mean and state covariance matrix in~\eqref{max_sigma} and~\eqref{contr_param}, are calculated for the worst-case disturbance distributions returned by the adversary policy $\gamma^*$. Since the separation principle holds, this implies that state estimation must be performed with the disturbance distribution $\Pdist_{w,t}^*$ fixed for all time stages.
\end{remark}

It follows from \Cref{cor:separ_se_under_wc} that the DRSE approach from \Cref{sec:DRSE} is ideal for state estimation.
If the prior state distributions are computed for the worst-case distributions $\Pdist_{w,t}^*, \; t=0,\dots,T-1$, then the optimal state estimator $\phi^*$, which solves the DRSE problem recursively, can be used.

\subsection{Solution to the DRSE problem} \label{sec:sol_DRSE}

We begin by examining a single instance of the DRSE problem~\eqref{eqn:DRSE} for a fixed time $t$, which constitutes a weighted DR-MMSE problem. 
For simplicity, we consider the case where the nominal distributions of initial state and measurement noise  are Gaussian, while the true distributions may not be Gaussian.\footnote{Our method remains valid for non-Gaussian nominal distributions, as detailed in~\Cref{app:GelbrichMMSE}.} 

\begin{assumption}\label{assump:Gauss}
    The nominal distributions for initial state $x_0$ and measurement noise $v_t$ for all $t$ are Gaussian, i.e., $\Qhat_{x,0}=\Gauss(\xnom, \Xnom)$ and $\Qhat_{v,t}=\Gauss(\hat{v}_t,\hat{\Sigma}_{v,t})$ with mean vectors $\xnom\in\real{n_x}, \hat{v}_t\in\real{n_y}$ and covariances matrices $\Xnom \in \psd{n_x}, \hat{\Sigma}_{v,t} \in \pd{n_y}$, respectively.
\end{assumption}

With this assumption, the DRSE problem can be reformulated into a finite convex program as follows.

\begin{lemma}
\label{lem:GelbrichMMSE} 
Suppose~\Cref{assump:Gauss} holds. Then, at the initial time stage, the state estimation problem~\eqref{eqn:DRSE} is equivalent to the following convex problem:
\begin{equation} \label{eqn:DRMMSE_init_opt}
\begin{split}
\max_{\Xpriorinit, \Sigma_{v,0}} \; & \Tr[\Theta_0(\Xpriorinit -\Xpriorinit C^{\top}  (C\Xpriorinit C^{\top} + \Sigma_{v,0})^{-1} C \Xpriorinit) ] \\
\mbox{s.t.} \; & \mathrm{Tr}[\Sigma_{v,0} + \hat{\Sigma}_{v,0} -2 \big(\hat{\Sigma}_{v,0}^\frac{1}{2} \Sigma_{v,0} \hat{\Sigma}_{v,0}^\frac{1}{2}\big)^{\frac{1}{2}}] \leq \theta_{v}^2, \\
&\mathrm{Tr}[\Xpriorinit + \Xnom -2 ((\Xnom)^{\frac{1}{2}} \Xpriorinit (\Xnom)^{\frac{1}{2}})^{\frac{1}{2}}] \leq \theta_{x_0}^2\\
&\Xpriorinit\in\psd{n_x}, \Sigma_{v,0}\in\pd{n_y}.
\end{split}
\end{equation}
Also, at any time stage $t =1,\dots, T-1$, the state estimation problem~\eqref{eqn:DRSE} for fixed prior state distribution $\Priordist = \Gauss(\xprior_t, \Xprior)$ is equivalent to the following convex problem:
\begin{equation}\label{eqn:DRMMSE_opt}
\begin{split}
\max_{\Sigma_{v,t}} \; & \Tr[\Theta_t(\Xprior-\Xprior C^{\top}  (C\Xprior C^{\top} + \Sigma_{v,t})^{-1} C \Xprior) ] \\
\mbox{s.t.} \; & \Tr[\Sigma_{v,t} + \hat{\Sigma}_{v,t} -2 \big(\hat{\Sigma}_{v,t}^\frac{1}{2} \Sigma_{v,t} \hat{\Sigma}_{v,t}^\frac{1}{2}\big)^{\frac{1}{2}}] \leq \theta_{v}^2\\
&\Sigma_{v,t}\in\pd{n_y}.
\end{split}
\end{equation}
In addition, if $(\Xpriorinitopt,\Sigma_{v,0}^{*})$ is the maximizing pair of~\eqref{eqn:DRMMSE_init_opt} and $\Sigma_{v,t}^{*}$ is the maximizer of~\eqref{eqn:DRMMSE_opt} at time $t$, then the maximum in~\eqref{eqn:DRSE} is attained by the Gaussian distributions $\Priordistinitopt = \Gauss(\xnom, \Xpriorinitopt)$ and $\Pdist_{v,t}^{*} = \Gauss(\hat{v}_t, \Sigma_{v,t}^{*})$ for each $t=0,\dots, T-1$.
Moreover, at any time stage $t$, the following affine estimator achieves the minimum in the state estimation problem~\eqref{eqn:DRSE}:
\begin{equation}\label{eqn:DR_estimator}
\begin{split}
\phi_{t}^*(y_{t}) =\Xprior C^{\top} (C\Xprior  C^{\top} + \Sigma_{v,t}^{*})^{-1}(y_{t}-C\xprior_t -\hat{v}_{t}) +\xprior_t,
\end{split}
\end{equation}
where $\Xpriorinit = \Xpriorinitopt$ and $\xprior_0 = \xnom$.
\end{lemma}
\Cref{lem:GelbrichMMSE} can be viewed as the adaptation of~\cite[Theorem 3.1]{nguyen2023bridging} to suit our particular context, where we have added a weighting matrix $\Theta_t$ to the cost and disregarded the ambiguities in the prior state distribution at $t>0$ due to our control structure.
Given the DR state estimator for a fixed time $t$ and a Gaussian prior state distribution $\Priordist$, we show that its iterative application leads to a DR Kalman filter. This enables the derivation of state estimates from new measurements. 

\begin{theorem}[DR Kalman Filter]\label{thm:DRKF}
    Suppose~\Cref{assump:Gauss} holds. Consider the system~\eqref{eqn:system eq} with control inputs $u_t^*$ and disturbances $w_t$ governed by the worst-case Gaussian distribution $\Pdist_{w,t}^* =\Gauss(\bar{w}_t^*, \Sigma_{w,t}^{*})$ at each time $t$. 
    Then, the DR state estimates $\phi^*_t(y_t)$, which recursively solve~\eqref{eqn:DRSE}, correspond to the conditional expectation $\xpost_t$ of the states. 
    Additionally, for each time stage $t$, the worst-case prior and posterior state distributions retain Gaussian forms, $\Priordistopt = \Gauss(\xprior_t, \Xprior)$ and $\Postdistopt = \Gauss(\xpost_t, \Xpost)$, respectively. These distributions are recursively computed for $t=0,\dots, T-1$ as follows:

\textbf{(Measurement Update)} Update $\xpost_t$ and $\Xpost$ based on the measurement $y_t$ as follows:
\begin{align}
\xpost_t &= \Xprior C^{\top} (C \Xprior C^{\top} + \Sigma_{v,t}^{*})^{-1} (y_{t}-C \xprior_t -\hat{v}_{t}) + \xprior_t \label{cond_mean}\\
\Xpost &= \Xprior-\Xprior C^{\top} (C { \Xprior } C^{\top} + \Sigma_{v,t}^{*})^{-1}   C \Xprior,\label{cond_cov}
\end{align}
where $\Sigma_{v,t}^{*}$ is the maximizer of~\eqref{eqn:DRMMSE_opt} for $t=1,\dots, T-1$, with initial values $\xprior_0 = \xnom, \Xpriorinit = \Xpriorinitopt$. Here $(\Xpriorinitopt, \Sigma_{v,0}^{*})$ is derived as the maximizer pair of~\eqref{eqn:DRMMSE_init_opt} .

\textbf{(State Prediction)} Predict $\xprior_{t+1}$ and $\Xpriornext$ as follows:
\begin{align}
    \begin{split}
        \xprior_{t+1} &= A \xpost_{t}  + B u_t^* + \bar{w}_t^* \label{cond_mean_prior}
    \end{split} \\
    \Xpriornext &= A \Xpost A^\top + \Sigma_{w,t}^{*},\label{cond_cov_prior}
\end{align}
with $\Sigma_{w,t}^{*}$ found as the maximizer of~\eqref{max_sigma} for $t=0,\dots,T-1$.
\end{theorem}

The proof of this theorem can be found in~\Cref{app:DRKF}.
Notably, the  measurement update and state prediction equations~\eqref{cond_mean}--\eqref{cond_cov_prior} mirror those of the standard Kalman filter when the system is assumed to be influenced by the worst-case distributions $\Pdist_{w,t}^*,\Pdist_{v,t}^{*}$, and $\Priordistinitopt$. Our DR Kalman filter differs from the one presented in~\cite{NEURIPS_DRKF}, which addresses the DRSE problem for the joint prior state and measurement distributions. 
In contrast, we we explicitly employ the worst-case distributions $\Pdist_{e,t}^*$ and $\Pdist_{w,t}^*$ along with the control inputs, fully exploiting the system structure.

\subsection{Unified SDP and algorithm for DRC and DRSE}\label{sec:integ}

\begin{algorithm}[t]
\DontPrintSemicolon
\SetKw{Input}{Input:}
\SetKw{to}{to}
\SetKw{and}{and}
\SetKw{break}{break}
\Input $Q_f,Q, R,  \lambda, \theta_v,  \theta_{x_0} , \Qhat_{x,0}, {\Qhat}_{w,t}, {\Qhat}_{v,t}$\;
\tcp{Offline Stage}
\For{$t=T-1,\dots, 0$}{
Compute $K_t, L_t, H_t, G_t$ according to~\eqref{KL} and~\eqref{HG}\; 
}
{Solve~\eqref{eqn:DRKF_SDP_init} in~\Cref{app:DRKF3} to obtain $(\Xpriorinitopt$, $\Xpostinit^*, \Sigma_{v,0}^{*})$ for initialization}\;
\For{$t=0,\dots, T-1$}{
Solve~\eqref{eqn:DRKF_SDP} forward in time to obtain;\; 
}
\tcp{Online Stage}
Observe $y_0$ and set $I_0 = y_0$\;
Estimate $\xpost_0$ via \eqref{eqn:DR_estimator}\;
\For{$t=0,\dots, T-1$}{
Apply $u_t^*=\pi_t^*(I_t)$ computed according to~\eqref{contr_param}\;
Compute the worst-case mean $\bar{w}_{t}^*$ using \eqref{mu}\;
Estimate $\xprior_{t+1}$ according to~\eqref{cond_mean_prior}\;
Observe $y_{t+1}$ and set $I_{t+1} = (I_{t}, y_{t+1}, u_{t}^*)$\;
Update $\xpost_{t+1}$
according to~\eqref{cond_mean}\;
}
\caption{WDR-CE} \label{alg:WDR-CE}
\end{algorithm}

In this section, we combine the DRC and DRSE components into a comprehensive WDR-CE algorithm, as outlined in \Cref{alg:WDR-CE}.
Recall that in~\Cref{thm:DRKF}, we assume the solvability of the problem in~\eqref{max_sigma}, which inherently requires a state estimator. To address this, we propose a novel reformulation of this problem into a tractable SDP problem by employing the DR Kalman filter. This strategy not only allows the computation of the worst-case covariance matrix $\Sigma_{w,t}^{*}$ but also integrates it with the DR state estimator, simplifying the overall computation process.

\begin{proposition}\label{prop:DRKF_SDP} 
Suppose~\Cref{assump:lambda,assump:Gauss} hold and the DR Kalman filter algorithm in~\Cref{thm:DRKF} is used for state estimation with $\Theta_{t+1} = S_{t+1}$. Then, $z_t(I_t)$ in~\eqref{max_sigma} corresponds to the optimal value of the following tractable SDP problem:
\begin{align}\label{eqn:DRKF_SDP}
    \begin{split}
    \max_{\substack{\Xpriornext,\Xpostnext \\\Sigma_{w,t}, \Sigma_{v,t+1}, Y, Z}} \; & \mathrm{Tr}[S_{t+1} \Xpostnext + (P_{t+1}-\lambda I)\Sigma_{w,t} + 2\lambda Y]\\
    \mbox{s.t.} \; 
    & \begin{bmatrix} \Xpriornext - \Xpostnext & \Xpriornext C^{\top} \\ C \Xpriornext & C \Xpriornext C^{\top} + \Sigma_{v,t+1} \end{bmatrix} \succeq 0\\
    & \begin{bmatrix}  \hat{\Sigma}_{w,t} & Y \\ Y^{\top} & {\Sigma}_{w,t} \end{bmatrix} \succeq 0,  \; 
    \begin{bmatrix} \hat{\Sigma}_{v,t+1} & Z \\ Z^{\top} & \Sigma_{v,t+1} \end{bmatrix} \succeq 0 \\
    &  \Xpriornext = A\Xpost A^{\top}+\Sigma_{w,t} \\
    & \mathrm{Tr}[\Sigma_{v,t+1} + \hat{\Sigma}_{v,t+1} -2 Z] \leq \theta_{v}^2 \\
    &  \Xpriornext, \Xpostnext, \Sigma_{w,t} \in\psd{n_x},\; \Sigma_{v,t+1} \in\pd{n_y} \\
    &Y\in\real{n_x\times n_x},\; Z \in\real{n_y\times n_y},
    \end{split}
\end{align}
where $\Xpost$ is the posterior covariance matrix of state $x_t$ conditioned on $I_t$.
Furthermore, let ($\Xpriornextopt, \Xpostnextopt,  \Sigma_{w,t}^{*},\Sigma_{v,t+1}^{*}, Y^*, Z^*$) be an optimal solution of \eqref{eqn:DRKF_SDP}. Then, $\Sigma_{v,t+1}^{*}$ is also the maximizer of~\eqref{eqn:DRMMSE_opt} at $t+1$, while $\Xpostnextopt$ and $\Xpriornextopt$ correspond to the posterior and prior state covariance matrices found according to~\eqref{cond_cov} and~\eqref{cond_cov_prior}, respectively. 
\end{proposition}

The proof of this proposition and the reformulation of \eqref{eqn:DRMMSE_init_opt} into a tractable SDP problem can be found in \Cref{app:DRKF3}.
The SDP problem~\eqref{eqn:DRKF_SDP} is independent of the measurements, allowing the matrices $\Xprior, \Xpost, \Sigma_{v,t}^{*}$, and $\Sigma_{w,t}^{*}$ to be predetermined offline for each time stage $t$.\footnote{Interestingly, if we ignore the ambiguities in the initial state and measurement noise by setting the ambiguity set radii $\theta_{v}$ and $\theta_{x_0}$ to zero, the SDP problem~\eqref{eqn:DRKF_SDP} reduces to the one in~\cite[Proposition 1]{hakobyan2024wasserstein}. This connection demonstrates the flexibility of our approach to handle various scenarios of distributional uncertainty.}
The choice of weighting matrices $\Theta_t = S_{t}$ is necessary for formulating the tractable SDP problem in~\Cref{prop:DRKF_SDP}. However, this choice is driven by the structure of the optimal value of the approximate DRC problem, where the estimation error influences the optimal value~\eqref{opt_val} through $S_t$. Consequently, in our DRSE problem, we aim to iteratively minimize these errors for the worst-case uncertainty distributions.

\begin{remark}
    The computational complexity of our algorithm primarily arises from solving the SDP problem~\eqref{eqn:DRKF_SDP}. While both our method and the approach in~\cite{taskesen2024distributionally} exhibit polynomial complexity with respect to problem size, our method solves distinct, smaller SDP problems over the time horizon $T$, resulting in slower growth in complexity.
    Specifically, when using an interior-point method~\cite{nemirovski2004interior}, the time complexity for finding all the worst-case covariance matrices with our approach is $\mathcal{O}(T(n_x^2 + n_y^2)^{3.5})$, compared to $\mathcal{O}((T(n_x^2 + n_y^2))^{3.5})$ for the SDP problem in~\cite{taskesen2024distributionally}.\footnote{Although~\cite{taskesen2024distributionally} introduces a Frank-Wolfe algorithm for efficiently solving the SDP problem, the linearization oracle induces numerical instabilities for long horizons.}
\end{remark}

Overall, our algorithm encompasses two main stages: \emph{offline} and \emph{online}. During the offline stage, all parameters for control and estimation, including the worst-case covariance matrices are precomputed. Then, in the online stage, the system implements the DR control policy $\pi_t^*$ and computes the mean of the worst-case disturbances $\bar{w}_t^*$, utilizing the real-time observations and the DR state estimator. 
This online adaptation of $\bar{w}_t^*$ using \eqref{mu}
is a unique feature that distinguishes our approach from \cite{taskesen2024distributionally}, where static zero-mean distributions for all uncertainty distributions are assumed.

\section{Performance guarantees}\label{sec:properties}
While our WDR-CE algorithm is both tractable and scalable, it is essential to assess its theoretical performance. This section begins by establishing the guaranteed cost property for our control policy, followed by an analysis of the out-of-sample performance of the overall WDR-CE algorithm.

\subsection{Guaranteed cost property}

The approximate DRC problem in~\Cref{sec:approxDRC} is essential for tractable solutions. However, a pivotal question arises: \emph{does the optimal control policy $\pi^*$ from this approximation maintain distributional robustness in the original DRC problem?}
The following proposition establishes the guaranteed cost property of our control policy $\pi^*$ for any worst-case distribution within the ambiguity set.

\begin{proposition}\label{prop:guarant_cost}
    Suppose~\Cref{assump:lambda} holds and the maximization problem in~\eqref{max_sigma} attains an optimal solution. Then, for any policy $\gamma\in\Gamma_{\mathcal{D}}$, we have that
\begin{equation}\label{eqn:J_c_upperbound}
J_c(\pi^*, \gamma, \phi^*, \Pdist_e^*)  \leq J_c^{\lambda}(\pi^*, \gamma^*, \phi^*, \Pdist_e^*) + \lambda \theta_w^2 T,
\end{equation}
where $J_c^{\lambda}(\pi^*, \gamma^*, \phi^*, \Pdist_e^*)$ represents the optimal value of the DRC problem when incorporating the DR Kalman filter with the worst-case distribution $\Pdist_e^*$.
\end{proposition}

Its proof can be found in~\Cref{app:guarant_cost}. This proposition highlights the distributional robustness of our controller, validating the approximation scheme. Besides, it provides a guideline for selecting the penalty parameter $\lambda$ based on the radius $\theta_w$. For instance, one can choose $\lambda$ to minimize the right-hand side of~\eqref{eqn:J_c_upperbound}, as detailed in \Cref{app:lambda_select}.

\subsection{Out-of-sample performance guarantee}

Suppose in our WDR-CE algorithm the nominal distributions are chosen as the empirical ones constructed from a finite sample dataset.\footnote{For example, given a dataset of i.i.d samples $\mathbf{w}_t=\{\hat{w}^{(i)}_t\}_{i=1}^{N}$, the empirical distribution of $w_t$ can be found as $\Qhat_{w,t} := \frac{1}{N} \sum_{i=1}^{N} \delta_{\hat{w}_t^{(i)}}$, with $\delta_w$ representing the Dirac measure concentrated at $w$.} Then, this raises the question: \emph{how well does the optimal control policy perform when tested with unseen  realizations of the uncertainties?}
It is known in the literature that using Wasserstein DRO with a well-calibrated ambiguity set can enhance out-of-sample performance and provide high-probability upper bounds~\cite{esfahani2015data, boskos2020data, gao2023finite}.
Inspired by this, we analyze the \textit{out-of-sample performance} of the controller and state estimator pair $(\pi_\mathbf{D}^*, \phi_\mathbf{D}^*)$ of our WDR-CE algorithm with the nominal distributions constructed from a training dataset $\mathbf{D} = \{\mathbf{w}, \mathbf{v},\mathbf{x_0}\}$, where $\mathbf{w}=\{\hat{w}_0^{(i)}, \dots, \hat{w}_{T-1}^{(i)}\}_{i=1}^{N}, \mathbf{v}=\{\hat{v}_0^{(i)}, \dots, \hat{v}_{T-1}^{(i)}\}_{i=1}^{N}$, and $\mathbf{x}_0=\{\hat{x}_0^{(i)}\}_{i=1}^{N}$ are $N$ i.i.d. realizations of the uncertainties, sampled from the true distributions.  We seek conditions on the ambiguity set radii for which the optimal value of the approximate DRC problem with dataset $\mathbf{D}$ provides an upper confidence bound on the out-of-sample performance.

\begin{theorem}\label{thm:osp}
Suppose that~\Cref{assump:lambda,assump:Gauss} hold. Also, suppose there exist constants $c_j > 2$ and $D_j > 0$ such that $\mathbb{E}_{d \sim \Qdist_{j,t}} [\exp (\|d\|^{c_j})] \leq D_{j}$ for $j\in \{w, v, x_0\}$. For the given confidence level $\beta \in (0,1)$, choose the radii $\theta_j$ for $j\in\{w, v, x_0\}$ according to
\begin{equation}\label{radius}
\theta_j := \begin{cases} a_j^{2/c_j} & \text{if}\; a_j > 1\\ a_j^{1/2} & \text{if}\; a_j \leq 1 \wedge n_j<4\\ a_j^{2 / n_j} & \text{if}\; a_j \leq 1 \wedge n_j>4\\ \bar{\theta}_j & \text{if}\; a_j \leq 1/(\log 3)^2 \wedge n_j=4,\end{cases}
\end{equation}
where $a_j:=(c_{2,j} N)^{-1}\log (c_{1,j} [1 - (1-\beta)^{1/(2T + 1)}]^{-1})$, 
and $\bar{\theta}_j$ satisfies the condition
$\bar{\theta}_j/\log (2+1/\bar{\theta}_j)=a_j^{1/2}$. Then, the following probabilistic out-of-sample performance guarantee holds:
\begin{equation}
\Qdist^{N}\{ \mathbf{D} \mid J_c(\pi^*_\mathbf{D},\gamma, \phi^*_\mathbf{D},\Qdist_e) \leq J_{c,\mathbf{D}}^{\lambda}(\pi^*_\mathbf{D}, \gamma^*_\mathbf{D}, \phi^*_\mathbf{D}, \Pdist^*_{e,\mathbf{D}}) +\lambda \theta_w^2 T\} \geq 1 - \beta.
\end{equation}

\end{theorem}
Its proof can be found in~\Cref{app:osp_combined}.
Although~\Cref{thm:osp} offers a theoretical method for selecting the radii, the ambiguity set it produces tends to be overly conservative and thus impractical. Some works in the DRO literature derive improved finite sample guarantees for the Wasserstein DRO~\cite{gao2023finite, blanchet2021sample, blanchet2022confidence, blanchet2019robust}. However, a more practical approach is to use cross-validation or bootstrapping techniques, which are prevalent in DRO literature, to determine the optimal radius~\cite{esfahani2015data, gao2022wasserstein, kannan2023residuals}.

\section{Numerical experiments}\label{sec:experi}

\begin{figure}[t!]
     \centering
     \begin{subfigure}{0.45\linewidth}
         \centering
         \includegraphics[trim={0 1cm 0 0.8cm},clip, width=\linewidth]{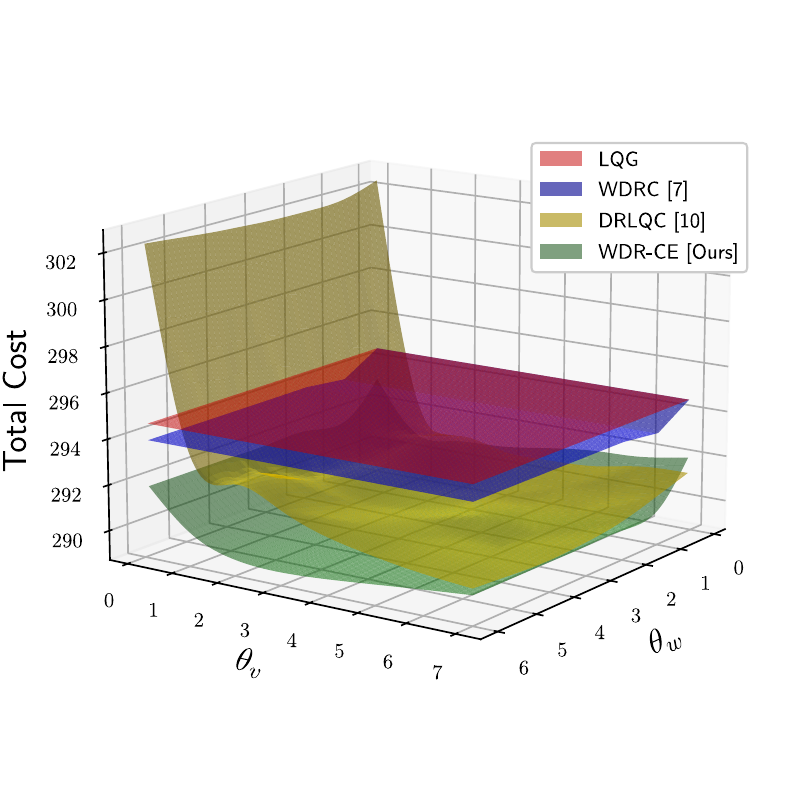}
    \caption{ }
    \label{fig:quad_quad_params_nonzeromean} 
     \end{subfigure}%
     \hspace{0.5in}
     \begin{subfigure}{0.45\linewidth}
         \centering
         \includegraphics[width=\linewidth]{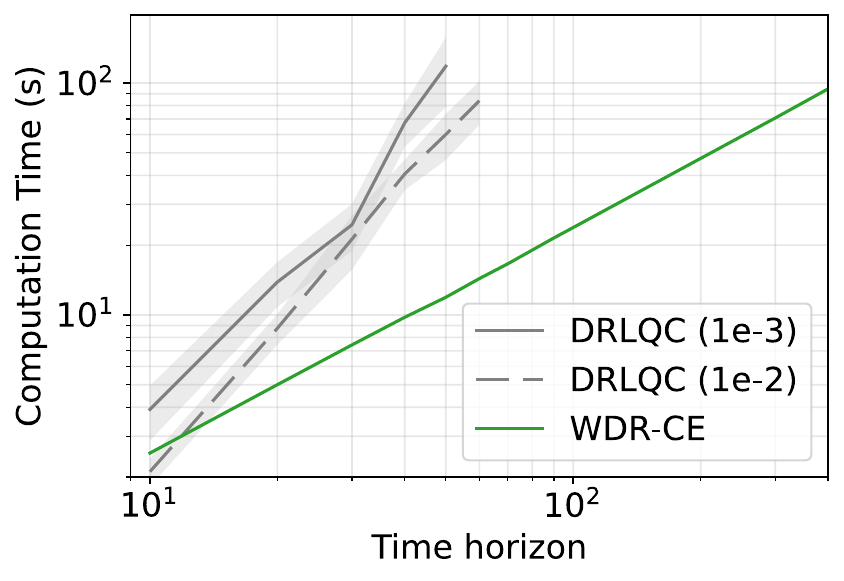}
    \caption{ }
    \label{fig:time_complexity} 
     \end{subfigure}
     \caption{(a) Total costs incurred by LQG, WDRC, DRLQC, and our WDR-CE methods, for nonzero-mean U-Quadaratic distribution with varying $\theta_w$ and $\theta_v$, averaged over 500 simulation runs; (b) Computation times for zero-mean Gaussian distributions averaged over 10 simulation runs. The shaded region represents 25\% standard deviation from the mean.}\label{fig:Time_and_Cost}
\end{figure}

We evaluate the WDR-CE scheme against three baselines: the LQG controller(e.g.,~\cite[Section 6.6.3]{kwakernaak1972linear}), the Wasserstein DRC (WDRC) method~\cite{hakobyan2024wasserstein}, and the DR-LQ control (DRLQC) method~\cite{taskesen2024distributionally}. Detailed experiment settings and additional results can be found in \Cref{app:experiment_details}.\footnote{Source code available at: \href{https://anonymous.4open.science/r/WDR-CE-D3CD}{\tt https://anonymous.4open.science/r/WDR-CE-D3CD}}

\Cref{fig:quad_quad_params_nonzeromean} shows the impact of ambiguity set sizes $\theta_w$ and $\theta_v$ on the total cost for uncertainties following a nonzero-mean U-Quadratic distribution. LQG performs the worst, while both DRLQC and WDR-CE show a decrease in total cost with increasing $\theta_v$ until an optimal point, after which the cost rises. Notably, WDR-CE consistently yields a lower total cost than DRLQC because DRLQC only accounts for uncertainties in the covariance matrix, neglecting potential errors in mean vectors. We evaluated scalability by comparing the computation times of WDR-CE and DRLQC for duality gap thresholds of $10^{-2}$ and $10^{-3}$ across different time horizons under zero-mean Gaussian uncertainties. \Cref{fig:time_complexity} shows that WDR-CE requires less computation time than DRLQC, regardless of the horizon. This superior scalability and efficiency make WDR-CE appealing for practical applications, especially with longer time horizons.

\section{Concluding remarks}\label{sec:concl}
We presented a novel WDR-CE algorithm that addresses distributional uncertainties in partially observable linear stochastic systems, where only limited nominal information is available. Our approach unifies DRC and DRSE problems in a novel way, proposing a tractable and scalable SDP formulation. The algorithm's several salient features are demonstrated in numerical experiments.
In future research, we plan to analyze the steady-state behavior and stability of the closed-loop system. 
It is also worth investigating an adaptive approach to systematically adjusting the sizes of ambiguity sets.

\appendix

\section{Additional details of \texorpdfstring{\Cref{sec:approxDRC}}{Theorem sol}}\label{app:sol}
To address the minimax problem~\eqref{eqn:minimax}, we first define the optimal cost-to-go in a  recursive manner as follows:
\begin{equation}\label{bellman}
\begin{split}
    V_t(I_t) \coloneqq \inf_{u_t \in \real{n_u}} \sup_{\Pdist_{w,t} \in \mathcal{P}(\real{n_x}) } \mathbb{E}_{x_t, y_{t+1}} \Bigl[ x_t^{\top}Qx_t  + u_t^{\top}Ru_t 
     &- \lambda G(\Pdist_{w,t}, \Qhat_{w,t})^2 \\
     &+ V_{t+1}(I_t, y_{t+1}, u_t) \mid I_t, u_t\Bigr] 
\end{split}
\end{equation}
for $t = T-1,\dots ,0$, with the terminal condition 
$V_T(I_T) = \mathbb{E}_{x_T}[x_T^\top Q_f x_T\mid I_T]$.
For analysis, assume the existence of an optimal solution $u_t^*$ of the outer minimization problem and $\Pdist_{w,t}^*$ of the inner maximization problem, as well as the measurability of the value function for all $t$. Consequently, for any state estimator and distribution pair $(\phi, \Qdist_e)$, the optimal cost for the approximate problem~\eqref{eqn:minimax} is determined by $J_c^{\lambda}(\pi^*,\gamma^*, \phi, \Qdist_e)= \mathbb{E}_{y_0}[V_0(I_0)]$, where $V_0$ is computed backward in time according to~\eqref{bellman}. 
Additionally, by setting $\gamma_t^{*}(I_t) = \Pdist_{w,t}^*$ and $\pi_t^*(I_t) = u_t^*$, an optimal policy pair $(\pi^*,\gamma^{*})$ can be obtained for the approximate problem~\eqref{eqn:minimax}. Therefore, it remains to solve the minimax problem on the right-hand side of~\eqref{bellman}, the solution of which is summarized in~\Cref{thm:sol} as per~\cite{hakobyan2024wasserstein}. 
The Riccati equation corresponding to \Cref{thm:sol} is given as follows:
\begin{align}
    P_t  =\, & Q + A^\top (I + P_{t+1}\Phi)^{-1} P_{t+1} A \label{P}\\
    S_t  = \, & Q + A^\top  P_{t+1} A - P_t\label{S}\\
    r_t  =  \, & A^\top (I + P_{t+1}\Phi)^{-1} ( r_{t+1} + P_{t+1}\hat{w}_{t})\label{r}\\
    \begin{split}
    q_t = \, & q_{t+1}  + (2\hat{w}_{t} - \Phi r_{t+1})^\top(I + P_{t+1}\Phi)^{-1} r_{t+1} +\hat{w}_{t}^\top (I + P_{t+1}\Phi)^{-1}  P_{t+1} \hat{w}_{t} - \lambda\mathrm{Tr}[\hat{\Sigma}_{w,t}]\label{q}
    \end{split}
\end{align}
with the terminal conditions $P_T = Q_f, S_T = 0, r_T = 0$, and $q_T = 0$.

\section{Comparison with other DR state estimators}\label{app:DRSE_compare}
Our DRSE formulation excludes the ambiguities in the prior state distributions for $t>0$. This design choice is driven by the construction of our DRC problem, which effectively manages distributional uncertainties arising from system disturbances, eliminating the need for additional conservativeness.

Specifically, let $(\phi_t^*, \Pdist_{e,t}^*)$ be the solution pair to our DRSE problem \eqref{eqn:DRSE} at time $t$ given a prior state distribution $\Priordist$. It is straightforward to verify that, at $t > 0$, the performance of our estimator is bounded by
\[
J_{e,t}(\phi_t^*, \Pdist_{v,t}^{*}, \Priordist) \leq \inf_{\phi_{t} \in \mathcal{F}_t} \sup_{\substack{\tilde{\Pdist}_{v,t} \in \ambset_{v,t} \\ \tilde{\Pdist}_{x,t}^{-} \in \ambset_{x,t}}} J_{e,t}(\phi_t, \tilde{\Pdist}_{v,t}, \tilde{\Pdist}_{x,t}^{-}),
\]
where $\ambset_{x,t}$ is an ambiguity set around the prior state distribution $\Priordist$. The right-hand side corresponds to the weighted version of the DR-MMSE estimator problem in~\cite{nguyen2023bridging}.

In contrast, the DR Kalman filter in~\cite{NEURIPS_DRKF} considers a single ambiguity set for the joint distribution of measurement noise and the state, which can be defined as
\[
\mathcal{D}_{xv,t}:= \left\{\tilde{\Pdist}_{xv,t}:=\tilde{\Pdist}_{x,t}^- \times \tilde{\Pdist}_{v,t} \in \mathcal{P}(\real{n_y} \times \real{n_x}) \mid W_2(\tilde{\Pdist}_{xv,t}, \Qhat_{xv,t}) \leq \theta, \Qhat_{xv,t}:=\Priordist\times \Qhat_{v,t} \right\}.
\]
Depending on the size of the chosen ambiguity set, this DR Kalman filter may result in more conservative estimators. For example, if $\theta:=\sqrt{\theta_v^2 + \theta_{x_0}^2}$, then the DR Kalman filter will be more conservative than the DR-MMSE estimator, i.e.,
\[
J_{e,t}(\phi_t^*, \Pdist_{v,t}^{*}, \Priordist) \leq \inf_{\phi_{t} \in \mathcal{F}_t} \sup_{\tilde{P}_{xv,t} \in \ambset_{xv,t}} J_{e,t}(\phi_t, \tilde{\Pdist}_{v,t}, \tilde{\Pdist}_{x,t}^{-}),
\]
where the right-hand side corresponds to the weighted version of the DR Kalman filter problem in~\cite{NEURIPS_DRKF}.
This conservativeness increases if the prior state distribution is already conservative, as is the case with our approach.
In~\Cref{sec:Trac_approx_and_sol}, we demonstrate that our controller requires state estimates computed for the worst-case system disturbance distribution identified in the DRC problem, which directly affects the prior state distribution.

\section{Application of WDR-CE to non-Gaussian distributions}\label{app:GelbrichMMSE}
In~\Cref{sec:sol_DRSE}, we demonstrated solving the DRSE problem for Gaussian nominal distributions, which can be extended to elliptical distributions with the same density-generating function. However, our method is also effective for non-elliptical cases.

Specifically, suppose $\Qhat_{x,0}$ and $\Qhat_{v,t}$ are arbitrary distributions with finite second-order moments, while $\Pdist_{w,t}^*$ is a worst-case distribution with mean vector $\bar{w}_t^*$ and covariance matrix $\Sigma_{w,t}^*$. Then, the DR Kalman filter in~\Cref{thm:DRKF} is valid and provides the best affine MMSE estimator under the worst-case uncertainty distributions $\Priordistinitopt$ and $\Pdist_{v,t}^*$. This is because $\phi_t^*$ solves the ordinary MMSE estimation problem for fixed uncertainty distribution.

Additionally, at each time $t$, consider the following Gelbrich ambiguity sets for measurement noise and initial state distributions:
\[
\mathcal{G}_{v,t} :=\left\{\Pdist_{v,t}\in\mathcal{P}(\real{n_y}) \mid \Gelb{\Pdist_{v,t}}{\Qhat_{v,t}} \leq \theta_v\right\}, \; \mathcal{G}_{x,0} :=\left\{\Priordistinit\in\mathcal{P}(\real{n_x}) \mid \Gelb{\Priordist}{\Qhat_{x,0}} \leq \theta_{x_0}\right\},
\]
respectively. Then, at each time $t$, the affine state estimator $\phi_t^*$ solves the following DRSE problem with the Gelbrich ambiguity sets:
\[
\min_{\phi_t \in \mathcal{A}_t} \max_{\Pdist_{e,t} \in \mathcal{G}_{e,t}} J_{e,t}(\phi_t, \Pdist_{v,t}, \Priordist),
\]
where $\mathcal{A}_t \subseteq \mathcal{F}_t$ denotes the space of all affine estimators, while the ambiguity set $\mathcal{G}_{e,t}$ is defined as $\mathcal{G}_{e,0} := \mathcal{G}_{v,0} \times \mathcal{G}_{x,0}$ and $\mathcal{G}_{e,t} := \mathcal{G}_{v,t}$  at $t=0$ and $t > 0$, respectively.
This follows directly from similar properties of the DR-MMSE estimation in~\cite[Proposition 5.1]{nguyen2023bridging}. Therefore, our WDR-CE algorithm with the DR Kalman filter in~\Cref{thm:DRKF} alongside the SDP problem~\eqref{eqn:DRKF_SDP} remains valid for non-Gaussian distributions.

\section{Penalty parameter selection}\label{app:lambda_select}

The choice of the penalty parameter for a given radius $\theta_w$ highly impacts the approximation quality and the resulting control performance. The guaranteed cost property in~\Cref{prop:guarant_cost} shows that the cost upper-bound depends on $\lambda$, allowing control of the distributional robustness by tuning $\lambda$. Indeed, it is desirable to select $\lambda$ to minimize the upper bound in \eqref{eqn:J_c_upperbound} as follows:
\begin{align}\label{lambda_opt}
    \lambda(\theta_w) \in \argmin_{\lambda > \hat{\lambda}} \left\{J_c^{\lambda}(\pi^*, \gamma^*, \phi^*, \Pdist_e^*) + \lambda \theta_w^2 T\right\},
\end{align}
where $\hat{\lambda}:= \inf\{\lambda \mid \lambda I \succ P_{t}, t=1,\dots,T\}$
represents the smallest value satisfying~\Cref{assump:lambda}. Since $\hat{\lambda}$ is the unique boundary point that separates the range of $\lambda$ by whether it satisfies~\Cref{assump:lambda}, as suggested in~\cite{kim2023distributional}, 
$\hat{\lambda}$ can be determined using binary search. In our partially observable setting, computing the worst-case cost requires solving an SDP program, unlike the fully observable case where the optimization problem on the right-hand side of~\eqref{lambda_opt} is a convex program. Nevertheless, the optimization problem for 
$\lambda$ can be solved using standard numerical solvers.

\section{Proofs}\label{sec:proofs}

\subsection{Proof of~\Cref{thm:DRKF}}\label{app:DRKF}
\begin{proof}
We proceed by induction to prove the theorem. For the base case, by applying~\Cref{lem:GelbrichMMSE}, we can reduce the DRSE problem to a finite convex program, as formulated in~\eqref{eqn:DRMMSE_init_opt}. This yields the DR state estimator $\phi_0^*(y_0)$ defined in~\eqref{eqn:DR_estimator}, with the worst-case prior state distribution $\Priordistinit=\Gauss(\xprior_0, \Xpriorinit)$ with $\xprior_0 = \xnom, \Xpriorinit = \Xpriorinitopt$, and the worst-case measurement noise distribution $\Pdist_{v,0}^{*} = \Gauss(\hat{v}_0, \Sigma_{v,0}^{*})$. Here, the pair $(\Xpriorinitopt, \Sigma_{v,0}^{*})$ is obtained as the solution to~\eqref{eqn:DRMMSE_init_opt}. Given the Gaussian nature of $\Priordistinit$ and $\Pdist_{v,0}^{*}$, it follows that the posterior distribution $\Postdistinit$ is also Gaussian, characterized by a mean vector $\xpost_0=\phi_0^*(y_0)$ found by~\eqref{cond_mean} and a covariance matrix $\Xpostinit$ determined by~\eqref{cond_cov}.
    
 For the induction step, assume the posterior state distribution at time $t-1$ is Gaussian, denoted as $\Pdist_{x,t-1} = \Gauss(\xpost_{t-1}, \Xpostprev)$. Then, under~\Cref{assump:Gauss},  
 it naturally follows that for the system~\eqref{eqn:system eq} governed by the DR optimal policy pair $(\pi^*, \gamma^*)$, the prior state distribution at time $t$ is also Gaussian, represented as $\Priordist = \Gauss(\xprior_{t}, \Xprior)$, where $\xprior_{t}$ and $\Xprior$ are computed using~\eqref{cond_mean_prior} and~\eqref{cond_cov_prior}, respectively. Upon observing $y_t$, the DRSE problem at time $t$ simplifies to a weighted DR-MMSE problem. As the prior distribution is Gaussian,~\Cref{lem:GelbrichMMSE} implies that the DR estimator can be computed by~\eqref{eqn:DR_estimator}, along with the worst-case measurement noise distribution $\Pdist_{v,t}^{*} = \Gauss(\hat{v}_t, \Sigma_{v,t}^{*})$, where $\Sigma_{v,t}^*$ is derived as the solution to the maximization problem~\eqref{eqn:DRMMSE_opt}.    
    Given that both $\Priordist$ and $\Pdist_{v,t}^{*}$ are Gaussian, the posterior distribution $\Postdist$ remains Gaussian, with its mean vector $\xpost_t$ matching the DR state estimate $\phi_t^*(y_t)$ computed in~\eqref{cond_mean} and covariance matrix $\Xpost$ found by~\eqref{cond_cov}.
    
    Therefore, we conclude that the DR Kalman filter algorithm iteratively solves the DRSE problem for all $t$, thereby completing the proof.
\end{proof}

\subsection{Proof of~\Cref{prop:DRKF_SDP}}\label{app:DRKF3}
\begin{proof}
First, it is straightforward that the DR Kalman filter equations~\eqref{cond_cov} and~\eqref{cond_cov_prior} can be integrated into the maximization problem~\eqref{max_sigma}.
However, even in that case, it is still challenging to solve the optimization problem, as it requires the worst-case measurement noise covariance matrix $\Sigma_{v,t+1}^{*}$. 
To address this, in the next step we show that, given $\Xpost$,~\eqref{max_sigma} can be reformulated into the following optimization problem:
\begin{equation}\label{max_sigma_ref}
\begin{split}
\max_{\substack{\Xpriornext,\Xpostnext\\ \Sigma_{w,t}, \Sigma_{v,t+1}}} \; & \mathrm{Tr}[S_{t+1} \Xpostnext  + (P_{t+1}-\lambda I)\Sigma_{w,t}+ 2\lambda \big(\hat{\Sigma}_{w,t}^\frac{1}{2} \Sigma_{w,t} \hat{\Sigma}_{w,t}^\frac{1}{2}\big)^{\frac{1}{2}}]   \\
\mbox{s.t.}  
\; & \Xpostnext=\Xpriornext-\Xpriornext C^{\top} (C\Xpriornext C^{\top} + \Sigma_{v,t+1})^{-1} C \Xpriornext  \\
&  \Xpriornext = A\Xpost A^{\top}+\Sigma_{w,t} \\
&  \mathrm{Tr}[\Sigma_{v,t+1} + \hat{\Sigma}_{v,t+1} -2 \big(\hat{\Sigma}_{v,t+1}^{\frac{1}{2}} \Sigma_{v,t+1} \hat{\Sigma}_{v,t+1}^{\frac{1}{2}}\big)^{\frac{1}{2}}] \leq \theta_{v}^2\\
&  \Xpriornext, \Xpostnext, \Sigma_{w,t} \in\psd{n_x},\; \Sigma_{v,t+1}\in\pd{n_y}.
\end{split}
\end{equation}
Suppose $(\Xpriornextopt, \Xpostnextopt, \Sigma_{w,t}^{*}, \Sigma_{v,t+1}^{*})$ is an optimal solution to~\eqref{max_sigma_ref}. Then, we need to verify that $\Sigma_{v,t+1}^{*}$  solves the optimization problem~\eqref{eqn:DRMMSE_opt} for $\Theta_{t+1} = S_{t+1}$. First, $\Sigma_{v,t+1}^{*}$ is feasible for~\eqref{eqn:DRMMSE_opt}.
Next, for every fixed $\Sigma_{w,t}\in\psd{n_x}$ and $\Xpriornext\in\psd{n_x}$, the covariance matrix $\Sigma_{v,t+1}^{*}$ maximizes $\Tr[S_{t+1} (\Xpriornext-\Xpriornext C^{\top} (C\Xpriornext C^{\top} + \Sigma_{v,t+1})^{-1} C \Xpriornext)]$.
This expression coincides with the objective of the DRSE problem with $\Theta_{t+1} = S_{t+1}$, thereby establishing $\Sigma_{v,t+1}^{*}$ as the optimal solution for the optimization problem~\eqref{eqn:DRMMSE_opt}.

Furthermore, leveraging the result from~\cite[Proposition 2]{malago2018wasserstein}, we have that
\[
\Tr\Big[\big(\hat{\Sigma}_{v,t+1}^{\frac{1}{2}}   \Sigma_{v,t+1}  \hat{\Sigma}_{v,t+1}^{\frac{1}{2}}\big)^{\frac{1}{2}}\Big] =\max \left\{\mathrm{Tr}[ Z ] \; \middle| \; Z \in \real{n_y\times n_y}, \begin{bmatrix} \hat{\Sigma}_{v,t+1} & Z \\ Z^{\top} & \Sigma_{v,t+1}  \end{bmatrix} \succeq 0\right\}
\]
and
\[
\Tr\Big[\big(\hat{\Sigma}_{w,t}^\frac{1}{2}   \Sigma_{w,t}  \hat{\Sigma}_{w,t}^\frac{1}{2}\big)^{\frac{1}{2}}\Big] =\max \left\{\mathrm{Tr}[ Y ] \; \middle| \; Y \in \real{n_x\times n_x}, \begin{bmatrix} \hat{\Sigma}_{w,t} & Y \\ Y^{\top} & \Sigma_{w,t}  \end{bmatrix} \succeq 0\right\}.
\]
Applying this reformulation to~\eqref{max_sigma_ref} and utilizing the standard Schur complement argument (e.g.,~\cite[Appendix 5.5]{boyd2004convex}) to the equality constraint in~\eqref{max_sigma_ref}, we derive the tractable SDP formulation presented in~\eqref{eqn:DRKF_SDP}.

Similarly, we can derive the SDP formulation for \eqref{eqn:DRMMSE_init_opt}, which addresses the DRSE problem at the initial time stage for $\Theta_0 = S_0$ as follow:
\begin{equation}\label{eqn:DRKF_SDP_init}
    \begin{split}
    \max_{\substack{\Xpriorinit,\Xpostinit \\ \Sigma_{v,0}, Y, Z}} \; & \mathrm{Tr}[ S_0\Xpostinit]\\
    \mbox{s.t.} \; 
    & \begin{bmatrix} \Xpriorinit - \Xpostinit & \Xpriorinit C^{\top} \\ C \Xpriorinit & C \Xpriorinit C^{\top} + \Sigma_{v,0} \end{bmatrix} \succeq 0\\
    & 
    \begin{bmatrix} \Xnom & Y \\ Y^{\top} & \Xpriorinit \end{bmatrix} \succeq 0 , \; \begin{bmatrix} \hat{\Sigma}_{v,0} & Z \\ Z^{\top} & \Sigma_{v,0} \end{bmatrix} \succeq 0 \\
    & \mathrm{Tr}[\Sigma_{v,0} + \hat{\Sigma}_{v,0} -2 Z] \leq \theta_{v}^2\\
    &\mathrm{Tr}[\Xpriorinit + \Xnom -2 Y] \leq \theta_{x_0}^2 \\
    &  \Xpriorinit, \Xpostinit \in\psd{n_x},\; \Sigma_{v,0} \in\pd{n_y}\\
    & Y\in\real{n_x\times n_x},\; Z \in\real{n_y \times n_y}.
    \end{split}
\end{equation}
Here, an optimal solution $(\Xpriorinitopt, \Sigma_{0}^{v,*})$ for \eqref{eqn:DRKF_SDP_init} also solves \eqref{eqn:DRMMSE_init_opt}, while $\Xpostinit^*$ represents the worst-case posterior state covariance matrix at the initial time.
\end{proof}
\subsection{Proof of \Cref{prop:guarant_cost}}\label{app:guarant_cost}
\begin{proof}
For any policy $\gamma \in \Gamma_\ambset$, we have the inequality
    \begin{equation}\label{guar_cost}
    \begin{split}
        J_c(\pi^*, \gamma, \phi^*, \Pdist_e^*) &\leq J_c^{\lambda} (\pi^*, \gamma, \phi^*, \Pdist_e^*) + \lambda \theta_w^2 T\\
        & \leq J_c^{\lambda}(\pi^*,\gamma^*, \phi^*, \Pdist_e^*) + \lambda \theta_w^2 T,
    \end{split}
    \end{equation}
where the first inequality holds because $\Pdist_{w,t} \in \ambset_{w,t}$ implies that $ \Gelb{\Pdist_{w,t}}{\Qhat_{w,t}} \leq \theta_w$, and the second inequality holds from the definition of the approximate DRC problem \eqref{eqn:minimax}.

\end{proof}
\subsection{Proof of~\Cref{thm:osp}}\label{app:osp_combined}
\begin{proof}

First, we observe that when both $\Qdist_{e,t}\in\ambset_{e,t}$ and $\Qdist_{w,t}\in\ambset_{w,t}$, the iterative application of the DR Kalman filter yields the following upper bound on the cost:
\begin{equation}
\begin{split}
J_c(\pi^*_\mathbf{D}, \gamma,\phi^*_\mathbf{D}, \Qdist_e) & \leq J_c^{\lambda}(\pi^*_\mathbf{D}, \gamma^*_\mathbf{D},\phi^*_\mathbf{D}, \Pdist_{e,\mathbf{D}}^{*}) + \lambda \theta_w^2 T.
\end{split}
\end{equation}

Next, suppose the assumptions of the theorem hold. Then, according to the measure concentration inequality for the Wasserstein distance presented in~\cite[Theorem 2]{fournier2015rate}, when the Wasserstein ambiguity set radius is chosen according to~\eqref{radius}, we obtain the following probabilistic bounds:
 \[
 \Qdist^{N}_{w,t}\left\{ \mathbf{w}_t  \mid \Qdist_{w,t} \in\ambset_{w,t} \right\}\geq (1-\beta)^{1/(2T + 1)}
 \]
 with similar bounds holding for $\mathbf{v}_t$ and $\mathbf{x_0}$.
Consequently, it follows that
\[
\begin{split}
    \Qdist^{N}\big\{ \mathbf{D} \mid  J_c(\pi^*_\mathbf{D},&\gamma,\phi^*_\mathbf{D}, \Qdist_e)  \leq \,J_c (\pi^*_\mathbf{D},\gamma^*_\mathbf{D},\phi^*_\mathbf{D}, \Pdist_{e,\mathbf{D}}^*) + \lambda \theta_w^2 T   \big\}\\
    \geq \, & \Qdist_{x,0}^{ N}\left\{ \mathbf{x}_0 \mid \Qdist_{x,0} \in \ambset_{x,0}\right\}\prod_{t=0}^{T-1}\Qdist_{w,t}^{N}\left\{ \mathbf{w}_t \mid \Qdist_{w,t} \in \ambset_{w,t}\right\}\Qdist_{v,t}^{N}\left\{ \mathbf{v}_t \mid \Qdist_{v,t} \in \ambset_{v,t}\right\} \\
    \geq \, & 1-\beta,
\end{split}
\]
which finalizes the proof.
\end{proof}

\section{Experiment details}\label{app:experiment_details}

\begin{figure}[t!]
     \centering
     \begin{subfigure}{0.49\linewidth}
         \centering
         \includegraphics[trim={0 1cm 0 0.8cm},clip, width=\linewidth]{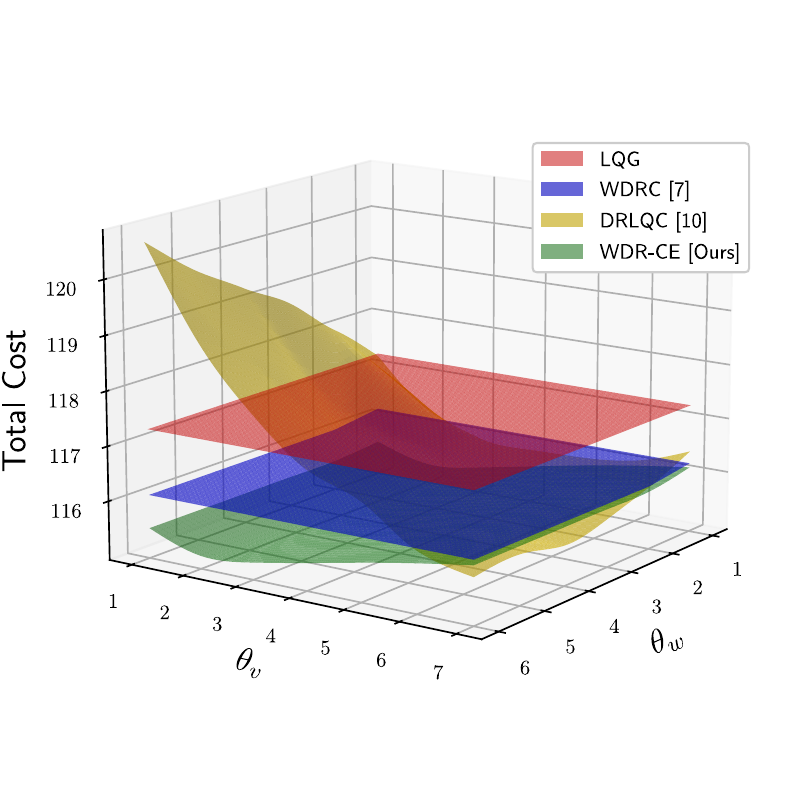}
    \caption{ }
    \label{appfig:norm_norm_params_zeromean} 
     \end{subfigure}%
     \begin{subfigure}{0.49\linewidth}
         \centering
         \includegraphics[trim={0 1cm 0 0.8cm},clip, width=\linewidth]{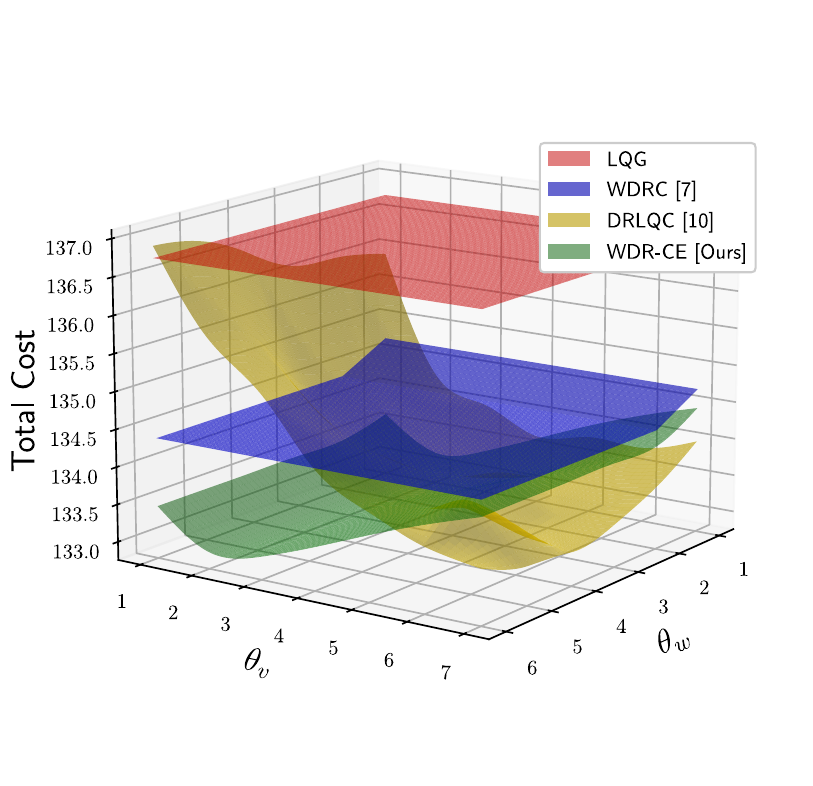}
    \caption{ }
    \label{appfig:quad_quad_params_zeromean} 
     \end{subfigure}
     \caption{Total costs incurred by LQG, WDRC, DRLQC, and our WDR-CE methods, for zero-mean (a) Gaussian and (b) U-Quadratic distributions, with varying $\theta_w$ and $\theta_v$, averaged over 500 simulation runs.}\label{appfig:hyperparam}
\end{figure}

All the numerical experiments were conducted on a computer equipped with an AMD Ryzen 7 3700X @ 3.60 GHz processor and 16GB RAM.

\subsection{Experiment settings for \Cref{sec:experi}}\label{app:exp_setting}

In these experiments, we consider the linear stochastic system described in~\cite{taskesen2024distributionally}. Specifically, we have a system with dimensions $n_x=n_u=n_y=10$, where the system matrix $A$ is defined element-wise as $A_{i,j} = 0.2$ if $i = j$ or $i = j-1$, and $A_{i,j} = 0$ otherwise. The other system matrices are set to $B=C=Q=Q_f=R = I_{10}$. The nominal distributions are constructed as empirical estimates with $N_{w} = N_{v} = N_{x_0}= 15$ samples for the system disturbance, measurement noise, and initial state distributions.

For comparison, we consider three baselines:
the standard LQG method (e.g.,~\cite[Section 6.6.3]{kwakernaak1972linear}), which uses the nominal distributions in both control and state estimation stages; the WDRC framework from~\cite{hakobyan2024wasserstein}, which solves the approximate DRC problem but uses the standard Kalman filter for state estimation; and the DRLQC method from~\cite{taskesen2024distributionally}, which solves the DRC problem for ambiguities in all uncertainty distributions but imposes a zero-mean Gaussianity assumption on the nominal distributions.
In the first experiment shown in~\Cref{fig:quad_quad_params_nonzeromean}, we compare the total costs for different ambiguity set sizes $\theta_w$ and $\theta_v$ with a fixed $\theta_{x_0} = 2$ and a horizon of $T=20$. Here, we consider nonzero-mean U-Quadratic distributions for all uncertainties: $w_t \sim \mathcal{UQ}(0, 2)$, $v_t \sim \mathcal{UQ}(-0.5, 2.5)$, and $x_0 \sim \mathcal{UQ}(0.8, 1.2)$. In the second experiment shown in~\Cref{fig:time_complexity}, we compare our method with the DRLQC framework in terms of scalability to time horizons, using zero-mean Gaussian distributions for all uncertainties: $w_t \sim \mathcal{N}(0, 0.1)$, $v_t \sim \mathcal{N}(0, 1.5)$, and $x_0 \sim \mathcal{N}(0, 0.1)$. The parameters for the ambiguity sets are set to $\theta_w = \theta_v = \theta_{x_0} = 1$.

\subsection{Additional experiments}\label{app:add_exp}
In addition to the experiments detailed in~\Cref{sec:experi}, we conducted further experiments in different scenarios to demonstrate the performance of our method.

\textbf{Zero-mean distributions.}
In this scenario, we compare the performance of our method for the same system as in~\Cref{app:exp_setting} to the baselines under two types of zero-mean probability distributions: Gaussian $(w_t \sim \mathcal{N}(0, 0.5), v_t \sim \mathcal{N}(0, 2), x_0 \sim \mathcal{N}(0, 0.1))$, and U-Quadratic $(w_t \sim \mathcal{UQ}(-1, 1), v_t \sim \mathcal{UQ}(-1.5, 1.5), x_0 \sim \mathcal{UQ}(-0.2, 0.2))$. Nominal distributions are constructed using $N =15$ samples. The time horizon is set to $T=20$, with the radius of the ambiguity set for the initial state fixed at $\theta_{x_0}=2$. The total costs for varying ambiguity set radii $\theta_w$ and $\theta_v$ are shown in~\Cref{appfig:norm_norm_params_zeromean} and~\Cref{appfig:quad_quad_params_zeromean}. The results demonstrate that the proposed algorithm performs effectively for both Gaussian and non-Gaussian distributions, outperforming the DRLQC for smaller $\theta_v$ values.

\textbf{Estimator performance.}
To demonstrate the efficacy of our DRSE estimator, we compare the proposed WDR-CE algorithm with three baselines based on the WDRC method, using: the standard Kalman filter for state estimation, the DR-MMSE estimator from~\cite{nguyen2023bridging} (WDRC+DRMMSE), and the DR Kalman Filter from~\cite{NEURIPS_DRKF} (WDRC+DRKF).

Comparisons are performed for two types of uncertainty distributions: Gaussian ($w_t \sim \Gauss(0.2,0.1)$, $v_t\sim \Gauss(0.2, 1.5)$, $x_0 \sim \Gauss(1,0.1)$) and U-Quadratic ($w_t \sim \UQ(1.0,-0.5)$, $v_t\sim \UQ(-1.5, 3.0)$, $x_0 \sim \UQ(0,0.5)$). We consider a system similar to that in~\Cref{app:exp_setting} with dimensions $n_x=n_u=10$ and $n_y=9$, and with system matrix $A$ structured element-wise as $A_{i,j} = 1$ if $i = j$ or $i = j-1$, and $A_{i,j} = 0$. The remaining system matrices are set to $B=Q=Q_f=R =I$ and $C=[I_9, 0_{9\times1}]$. The time horizon is set to $T=20$, and the ambiguity set radius for the initial state distribution is fixed at $\theta_{x_0}=5$. For DR Kalman filter~\cite{NEURIPS_DRKF}, we set $\theta=\sqrt{\theta_{x_0}^2 + \theta_v^2}$ as explained in~\Cref{app:DRSE_compare}. For the Gaussian distribution, we use $N_w=N_v=15$ and $N_{x_0}=10$ samples to construct nominal distributions, while for the U-Quadratic distribution, we use $N_w = N_v = N_{x_0}=20$ samples.

\begin{figure}[t!]
     \centering
     \begin{subfigure}{0.49\linewidth}
         \centering
         \includegraphics[trim={0 1cm 0 0.8cm},clip, width=\linewidth]{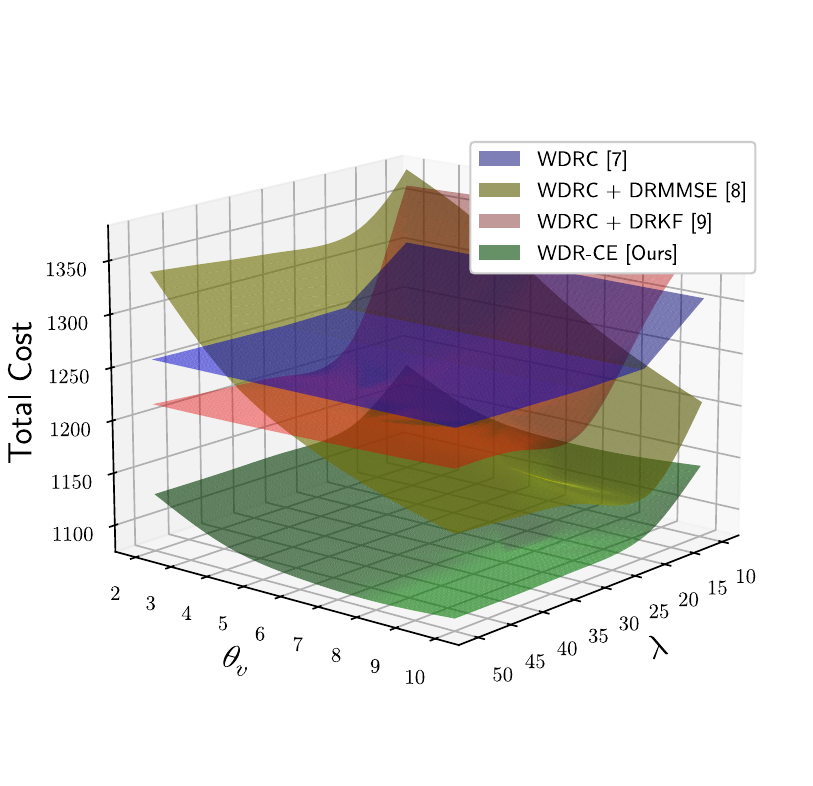}
    \caption{ }
    \label{fig:norm_norm_params9} 
     \end{subfigure}%
     \begin{subfigure}{0.49\linewidth}
         \centering
         \includegraphics[trim={0 1cm 0 0.8cm},clip, width=\linewidth]{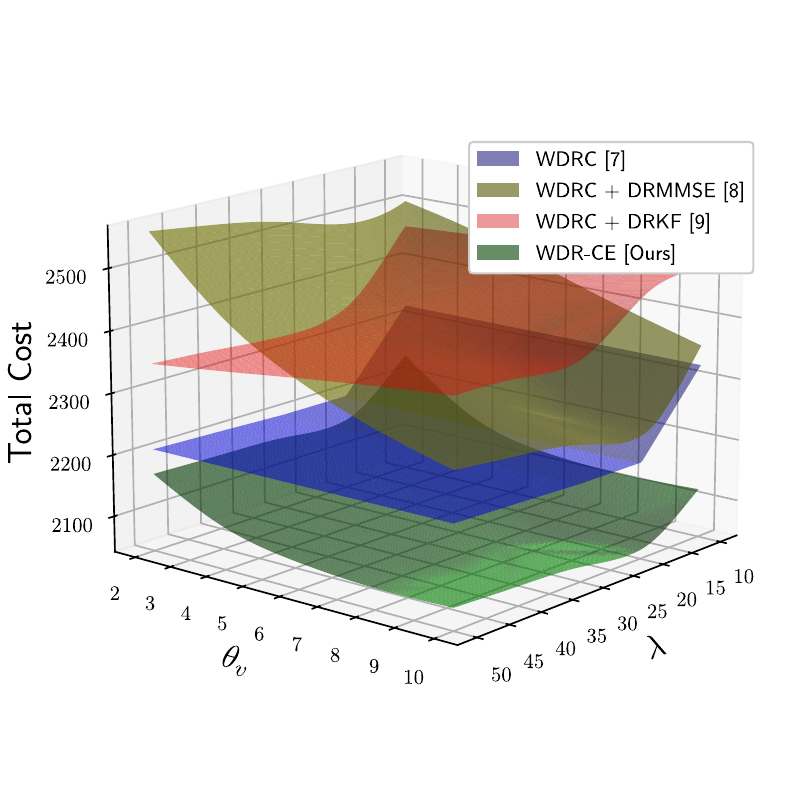}
    \caption{ }
    \label{fig:quad_quad_params9} 
     \end{subfigure}
     \caption{Total costs incurred by WDRC, WDRC+DRKF, WDRC+DRMMSE, and our WDR-CE methods, for (a) Gaussian and (b) U-Quadratic distributions, with varying $\lambda$ and $\theta_v$, averaged over 500 simulation runs.}\label{app:fig:Conserv_filter}
\end{figure}

In both scenarios, the only difference is the type of state estimator combined with the WDRC method. As shown in \Cref{fig:norm_norm_params9} and \Cref{fig:quad_quad_params9}, there is a clear difference in total cost among the four methods, highlighting the importance of selecting an appropriate state estimator. Notably, our WDR-CE method results in lower total costs compared to the other methods. This is due to the excessively large covariance estimates of WDRC+DRMMSE and WDRC+DRKF, which arise from the redundant ambiguity set with prior state distributions. Consequently, the overly conservative nature of these estimators degrades estimation quality and leads to higher costs. Interestingly, in the case of the U-Quadratic distribution, WDRC without any robust filter outperforms the other two baselines. This is because the WDRC method already effectively manages ambiguities from disturbance distributions, making additional robustness for prior state distributions unnecessary. 

\textbf{Problems with long time horizons.}
In this scenario, we compared our method with the LQG and WDRC methods over a long time horizon of $T=200$. Notably, we do not include the DRLQC method in this experiment due to its numerical instability for longer time horizons with non-convergent system matrices. 
We use the same system as in~\Cref{app:exp_setting} with a different system matrix, which is constructed element-wise as $A_{i,j} = 1$ if $i = j$ or $i = j-1$, and $A_{i,j} = 0$ otherwise. To evaluate the performance of our method, we explore two types of distributions for uncertainties: Gaussian ($w_t \sim \Gauss(0.1,0.1)$, $v_t\sim \Gauss(0.5, 2)$, $x_0 \sim \Gauss(0.1,0.1)$) and U-Quadratic ($w_t \sim \UQ(-0.5,0.2)$, $v_t \sim \UQ(-1,2)$, $x_0\sim \UQ(0,0.5)$). For both distributions, $N_{w} = N_{v} = N_{x_0}=15$ samples are used to construct nominal distributions.

\begin{figure}[t!]
     \centering
     \begin{subfigure}{0.49\linewidth}
         \centering
         \includegraphics[trim={0 1cm 0 0.8cm},clip, width=\linewidth]{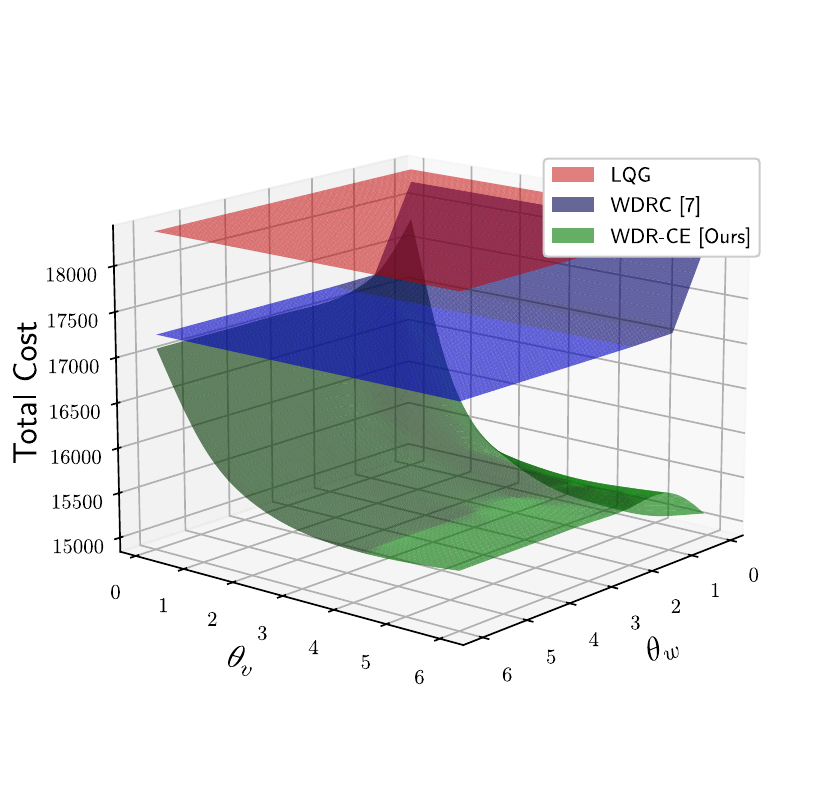}
    \caption{ }
    \label{fig:norm_norm_params_long} 
     \end{subfigure}%
     \begin{subfigure}{0.49\linewidth}
         \centering
         \includegraphics[trim={0 1cm 0 0.8cm},clip, width=\linewidth]{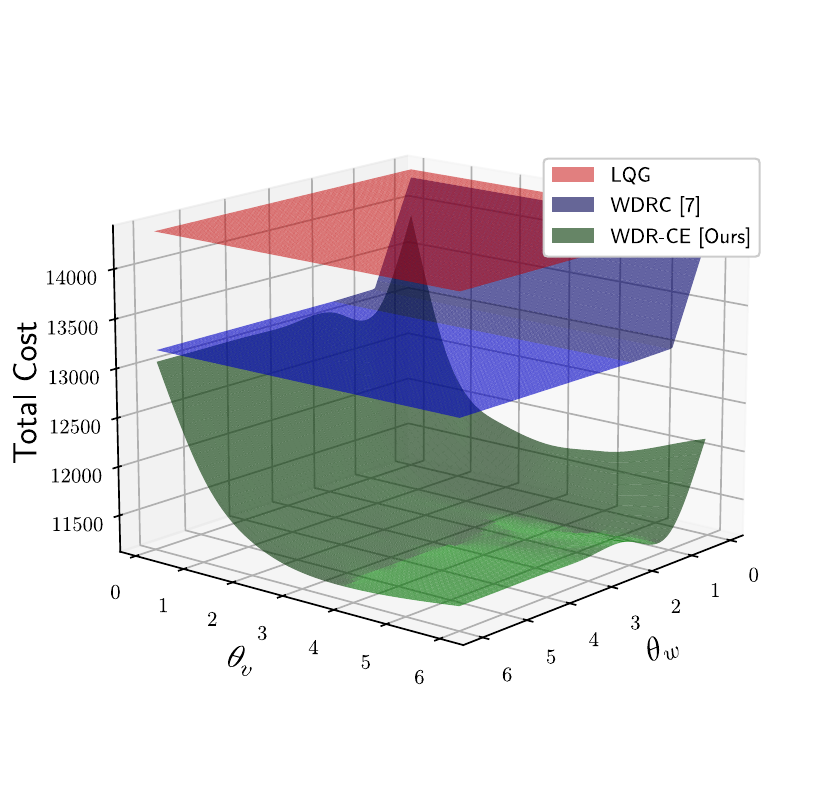}
    \caption{ }
    \label{fig:quad_quad_params_long} 
     \end{subfigure}
     \caption{Total costs incurred by LQG, WDRC, and our WDR-CE methods for a long horizon $T=200$, for (a) Gaussian and (b) U-Quadratic distributions, with varying $\theta_w$ and $\theta_v$, averaged over 500 simulation runs.}\label{app:fig:LongT}
\end{figure}

\Cref{fig:norm_norm_params_long} and \Cref{fig:quad_quad_params_long} illustrate the influence of the ambiguity set sizes $\theta_v$ and $\theta_w$ on the total costs for Gaussian distributions with $\theta_{x_0}=1$ and U-Quadratic distributions with $\theta_{x_0}=0.5$, respectively. In both scenarios, the WDR-CE method demonstrates a lower cost compared to the LQG and WDRC methods across the examined range of radii. Additionally, as $\theta_v$ approaches zero, the performance of WDR-CE converges to that of WDRC, as the DRSE converges to the standard Kalman Filter. 
This experiment, conducted over a long time horizon, demonstrates that our WDR-CE approach effectively handles scenarios with long time horizons when appropriate ambiguity set radii are used, highlighting its practical utility.

\begin{figure}[t!]
     \centering
     \begin{subfigure}{0.49\linewidth}
         \centering
         \includegraphics[width=\linewidth]{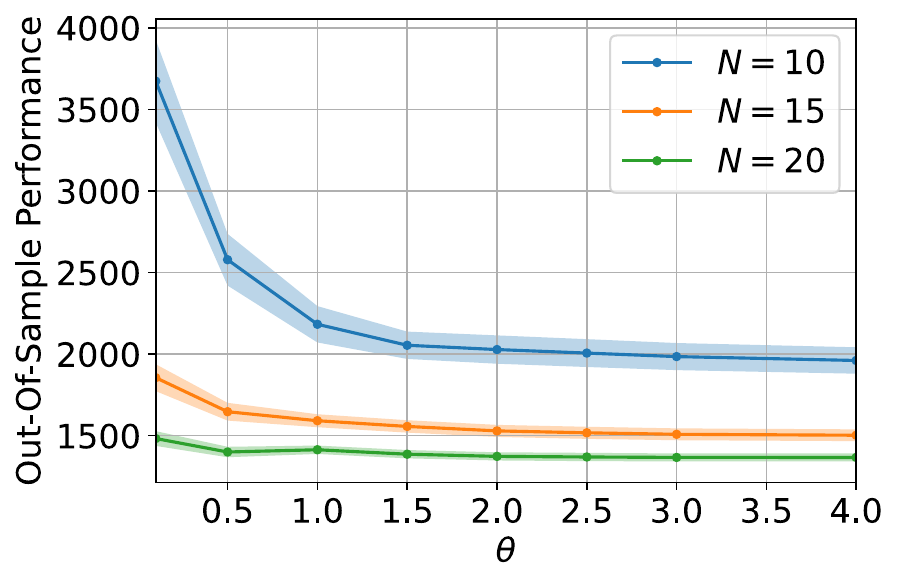}
    \caption{ }
    \label{fig:osp_normal} 
     \end{subfigure}%
     \begin{subfigure}{0.49\linewidth}
         \centering
         \includegraphics[width=0.97\linewidth]{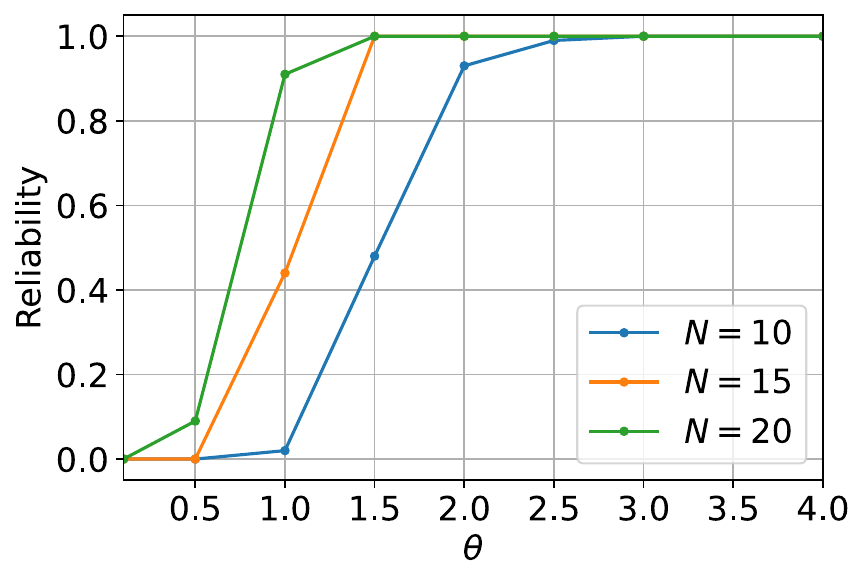}
    \caption{ }
    \label{fig:osp_prob_normal} 
     \end{subfigure}
     \caption{(a) Out-of-sample performance and (b) reliability of our WDR-CE method estimated over 100 simulation runs with 1000 uncertainty samples. The shaded region represents $25\%$ standard deviation from the mean.}\label{app:fig:OSP}
\end{figure}

\vspace{0.1in}
\textbf{Out-of-sample performance.}
In this scenario, we assess the out-of-sample performance of our method using the same system as in the previous experiment, with a time horizon of $T=20$. We consider Gaussian distributions for all the uncertainties: $w_t \sim \Gauss(0.1,0.1)$, $v_t\sim \Gauss(0.5, 2)$, $x_0 \sim \Gauss(0.1,0.1)$. The nominal distributions are constructed using varying sample sizes $N_{w} = N_{v} = N_{x_0} = N=\{10,15,20\}$.

\Cref{fig:osp_normal} shows the out-of-sample performance of our approach for varying radii $\theta_w = \theta_v = \theta_{x_0} = \theta$, computed over 100 independent simulation runs with 1000 uncertainty samples each. As the ambiguity set radius $\theta$ increases, the out-of-sample cost decreases, indicating better performance. Notably, when $N=10$, the out-of-sample performance is relatively poor for small $\theta$ values, as they are not sufficient to handle the distributional uncertainties in nominal distributions derived from small datasets. Additionally, as the sample size $N$ increases, the out-of-sample performance improves, as the nominal distributions become closer to the true distributions.

\Cref{fig:osp_prob_normal} shows the reliability of the WDR-CE algorithm for different sample sizes used to construct the nominal distributions, which is defined as the probability in~\eqref{eqn:J_c_upperbound}. As $\theta$ increases, the reliability of the algorithm increases to $1.0$, indicating that the true distributions are contained within the ambiguity sets. Additionally, as $N$ increases, reliability improves as the nominal distributions better approximate the true distributions.
This extensive assessment demonstrates the out-of-sample performance of the WDR-CE algorithm for unseen uncertainty realizations, making our algorithm useful in real-world applications.

\textbf{Vehicle control problem.}
In this scenario, we applied our method to a larger system that models position and velocity control for high-speed vehicles, as described in~\cite{leibfritz2006compleib}. 
The system is characterized by $n_x=21, n_u=11$, and $n_y=10$ states, control inputs, and outputs, respectively. The associated system matrices are configured with $Q=Q_f=I_{21}$ and $R=I_{11}$. The time horizon for the DRC problem~\eqref{eqn:minimaxbegin} is set to $T=20$. To evaluate the performance of our method, we explore two types of distributions for uncertainties: Gaussian ($w_t \sim \Gauss(1,0.1)$, $v_t\sim \Gauss(1, 0.5)$, $x_0 \sim \Gauss(0.1,0.1)$) and U-Quadratic ($w_t \sim \UQ(-0.4,0.8)$, $v_t \sim \UQ(-1,1.5)$, $x_0\sim \UQ(0,0.5)$).

\begin{figure}[t!]
     \centering
     \begin{subfigure}{0.49\linewidth}
         \centering
         \includegraphics[width=\linewidth]{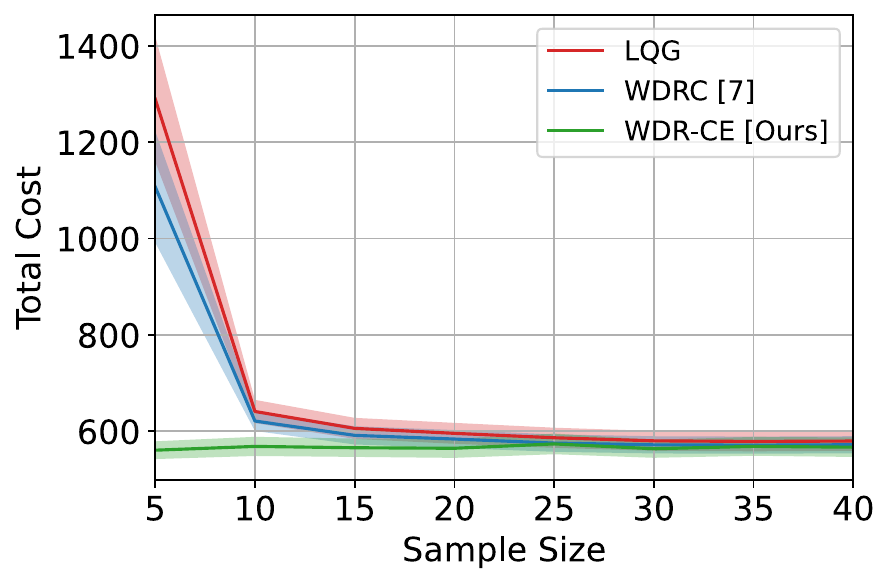}
    \caption{ }
    \label{fig:norm_norm_21_noise} 
     \end{subfigure}%
     \begin{subfigure}{0.49\linewidth}
         \centering
         \includegraphics[width=\linewidth]{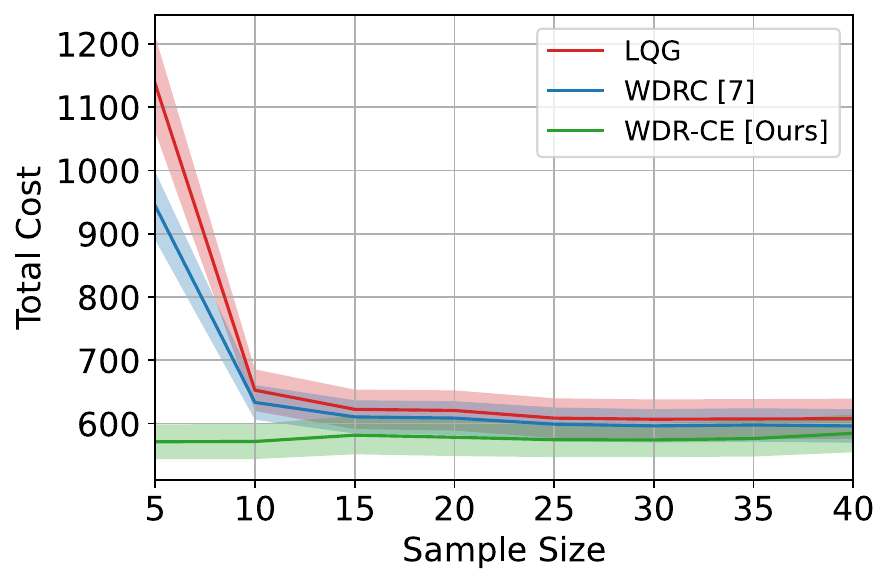}
    \caption{ }
    \label{fig:quad_quad_21_noise} 
     \end{subfigure}
     \caption{Total costs incurred by LQG, WDRC, and our WDR-CE methods for the vehicle control problem, for (a) Gaussian and (b) U-Quadratic distributions, with varying noise samples $N_v$, averaged over 500 simulation runs. The shaded region represents $25\%$ standard deviation from the mean.
     }\label{app:fig:nx=21_noiseplot}
\end{figure}
\begin{figure}[t!]
     \centering
     \begin{subfigure}{0.49\linewidth}
         \centering
         \includegraphics[trim={0 1cm 0 0.8cm},clip, width=\linewidth]{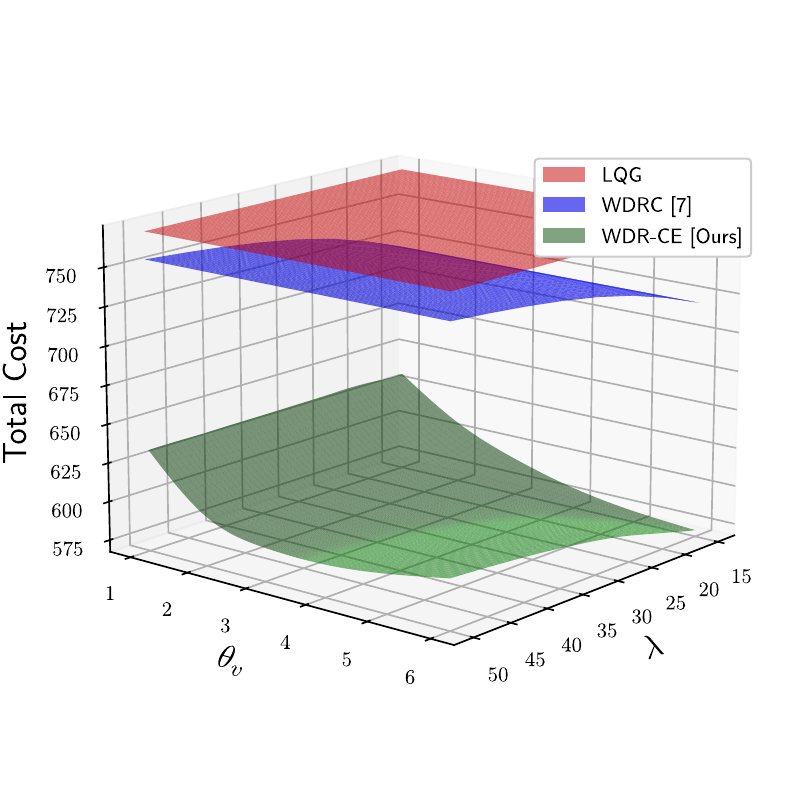}
    \caption{ }
    \label{fig:norm_norm_params21} 
     \end{subfigure}%
     \begin{subfigure}{0.49\linewidth}
         \centering
         \includegraphics[trim={0 1cm 0 0.8cm},clip, width=\linewidth]{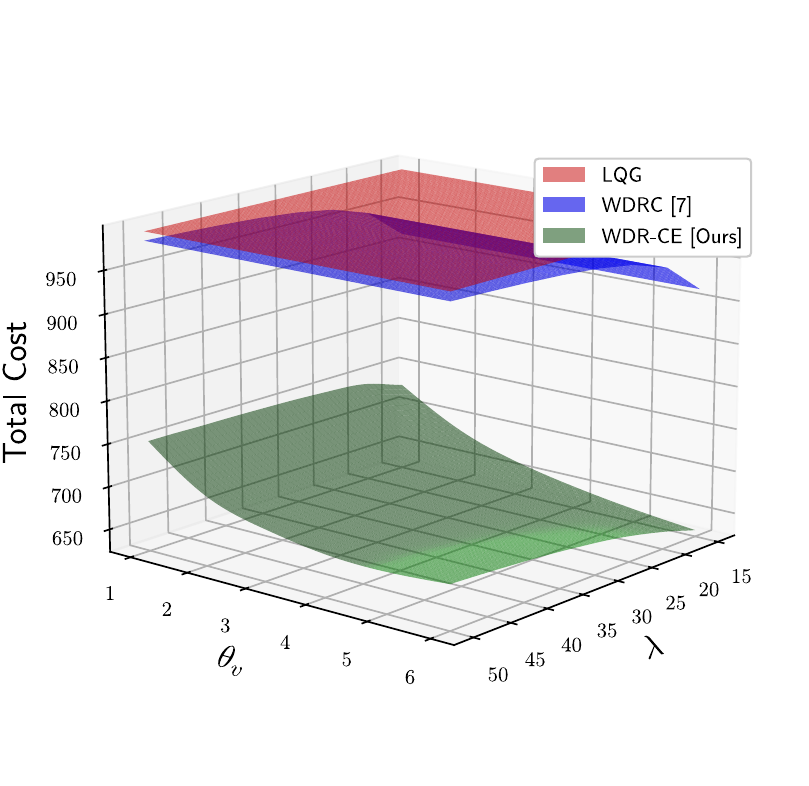}
    \caption{ }
    \label{fig:quad_quad_params21} 
     \end{subfigure}
     \caption{Total cost incurred by LQG, WDRC, and our WDR-CE methods for the vehicle control problem, for (a) Gaussian and (b) U-Quadratic distributions, for varying $\theta_w$ and $\theta_v$, averaged over 500 simulation runs.}\label{app:fig:nx=21_sys}
\end{figure}

\Cref{fig:norm_norm_21_noise} and~\Cref{fig:quad_quad_21_noise} show the total costs incurred by the LQG, WDRC, and WDR-CE methods for different sample sizes $N_v$ used for the nominal measurement noise distribution $\Qhat_t^v$, under Gaussian and U-Quadratic distributions, respectively. The sample sizes for system disturbances and initial state are kept constant at $N_w = N_{x_0} = 15$. For the Gaussian case, the ambiguity set parameters are $\theta_w=\theta_{x_0}=0.5$ and $\theta_{v}=5$, while for the U-Quadratic scenario, they are set to $\theta_w=1.0$, $\theta_{x_0}=0.5$ and $\theta_{v}=5$. 
The results indicate that the LQG approach tends to incur higher costs with smaller sample sizes, as it relies heavily on nominal distributions. Although the WDRC method does not explicitly address ambiguities in the measurement noise distribution, it manages them to an extent through the DR controller. However, the WDRC approach employs the standard Kalman filter, which is optimal only under exactly known Gaussian distributions. As the sample size $N_v$ increases, the average total cost decreases for both WDRC and LQG methods, as the nominal distribution becomes closer to the true one. In contrast, our WDR-CE method consistently outperforms both WDRC and LQG methods, demonstrating its efficacy even with small sample sizes.

\Cref{fig:norm_norm_params21} and \Cref{fig:quad_quad_params21} illustrate the effect of the penalty parameter $\lambda$ and radius $\theta_v$ on the total cost when $\theta_{x_0}=0.5$ and $N_w = N_v = N_{x_0} = 10$. The results show that WDR-CE consistently achieves a lower cost compared to LQG and WDRC, demonstrating the effectiveness of our method in practical systems.

\bibliographystyle{IEEEtran}

\medskip
\bibliography{ref}

\end{document}